\documentclass[11pt]{article}
\usepackage{graphicx}
\usepackage{amsmath}
\usepackage{amssymb}
\usepackage{latexsym}
\usepackage{bm}
\usepackage{overpic}
\usepackage[dvips]{color}
\usepackage{cases}
\usepackage{lscape}
\usepackage{here}
\usepackage{ulem}
\newcommand{\ds}{\displaystyle}
\newtheorem{theorem}{Theorem}
\newtheorem{lemma}{Lemma}
\newtheorem{remark}{Remark}
\newtheorem{example}{Example}

\begin{document}
%
%\linenumbers
%
\begin{center}
{\LARGE\bf Analysis for the Slow Convergence in Arimoto Algorithm\\[7mm]
\large Kenji~Nakagawa$^\ast$, Yoshinori~Takei$^\dagger$, Kohei~Watabe$^\ast$}
\end{center}
\footnote[0]{The material in this paper was presented in part at 2017 Symposium on Information Theory and its Applications (SITA2017).\\
$^\ast$Department of Electrical and Electronics and Information Engineering, Nagaoka University of Technology, Nagaoka, Niigata 940-2188, Japan, e-mail:nakagawa@nagaokaut.ac.jp, $\dagger$National Institute of Technology, Akita College, Akita 011-8511, Japan.}
\begin{abstract}
In this paper, we investigate the convergence speed of the Arimoto algorithm. By analyzing the Taylor expansion of the defining function of the Arimoto algorithm, we will clarify the conditions for the exponential or $1/N$ order convergence and calculate the convergence speed. We show that the convergence speed of the $1/N$ order is evaluated by the derivatives of the Kullback-Leibler divergence with respect to the input probabilities. The analysis for the convergence of the $1/N$ order is new in this paper. Based on the analysis, we will compare the convergence speed of the Arimoto algorithm with the theoretical values obtained in our theorems for several channel matrices.
\end{abstract}


Keywords:channel capacity, discrete memoryless channel, Arimoto algorithm, convergence speed, Hessian matrix.

\baselineskip 5.5mm

\section{Introduction}
Arimoto \cite{ari} proposed a sequential algorithm for calculating the channel capacity $C$ of a discrete memoryless channel. Based on the Bayes probability, the algorithm is given by the alternating minimization between the input probabilities and the reverse channel matrices. For arbitrary channel matrix $\Phi$ the convergence of the Arimoto algorithm is proved and the convergence speed is evaluated. In the worst case, the convergence speed is the $1/N$ order, and if the input distribution $\bm\lambda^\ast$ that achieves the channel capacity $C$ is in the interior of the set $\Delta({\cal X})$ of input distributions, the convergence is exponential. 

In this paper, we first consider the exponential convergence and evaluate the convergence speed. We show that there exist cases of exponential convergence even if $\bm\lambda^\ast$ is on the boundary of $\Delta({\cal X})$. Moreover, we also consider the convergence of the $1/N$ order, which is not dealt with in the previous studies. Especially, when the input alphabet size $m=3$, we will analyze the convergence of the $1/N$ order in detail and the convergence speed is evaluated by the derivatives of the Kullback-Leibler divergence with respect to the input probabilities.

As a basic idea for evaluating the convergence speed, we consider that the function $F(\bm\lambda)$ which defines the Arimoto algorithm is a differentiable mapping from $\Delta(\cal X)$ to $\Delta(\cal X)$, and notice that the capacity achieving input distribution $\bm\lambda^\ast$ is the fixed point of $F(\bm\lambda)$. Then, the convergence speed is evaluated by analyzing the Taylor expansion of $F(\bm\lambda)$ about the fixed point $\bm\lambda=\bm\lambda^\ast$.
\section{Related works}
There have been many related works on the Arimoto algorithm. For example, extension to different types of channels \cite{nai},\,\cite{rez},\,\cite{von}, acceleration of the Arimoto algorithm\,\cite{mat},\,\cite{yu}, characterization of Arimoto algorithm by divergence geometry \cite{csi2},\,\cite{mat},\,\cite{naj}, etc. If we focus on the analysis for the convergence speed of the Arimoto algorithm, we see in \cite{ari},\cite{mat},\cite{yu} that the eigenvalues of the Jacobian matrix are calculated and the convergence speed is investigated in the case that $\lambda^\ast$ is in the interior of $\Delta(\cal X)$.

In this paper, we consider the Taylor expansion of the defining function of the Arimoto algorithm. We will calculate not only the Jacobian matrix of the first order term of the Taylor expansion, but also the Hessian matrix of the second order term, and examine the convergence speed of the exponential or $1/N$ order based on the Jacobian and Hessian matrices. Because our approach for the evaluation of the convergence speed is very fundamental, we hope that our results will be applied to all the existing works.

\section{Channel matrix and channel capacity}
Consider a discrete memoryless channel $X\rightarrow Y$ with the input source $X$ and the output source $Y$. Let ${\cal X}=\{x_1,\cdots,x_m\}$ be the input alphabet and ${\cal Y}=\{y_1,\cdots,y_n\}$ be the output alphabet.

The conditional probability that the output symbol $y_j$ is received when the input symbol $x_i$ was transmitted is denoted by $P^i_j=P(Y=y_j|X=x_i),\,i=1,\cdots,m, j=1,\cdots,n,$ and the row vector $P^i$ is defined by $P^i=(P^i_1,\cdots,P^i_n),\,i=1,\cdots,m$. The channel matrix $\Phi$ is defined by\\[-3mm]
\begin{align}
\label{eqn:thechannelmatrix}
\Phi=\begin{pmatrix}
\,P^1\,\\
\vdots\\
\,P^m\,
\end{pmatrix}
=\begin{pmatrix}
\,P^1_1 & \cdots & P^1_n\,\\
\vdots & & \vdots\\
\,P^m_1 & \cdots & P^m_n\,
\end{pmatrix}.
\end{align}
We assume that for any $j\,(j=1,\cdots,n)$ there exist at least one $i\,(i=1,\cdots,m)$ with $P^i_j>0$. This means that there are no useless output symbols.

The set of input probability distributions on the input alphabet ${\cal X}$ is denoted by $\Delta({\cal X})\equiv\{\bm\lambda=(\lambda_1,\cdots,\lambda_m)|\lambda_i\geq0,i=1,\cdots,m,\sum_{i=1}^m\lambda_i=1\}$. The interior of $\Delta({\cal X})$ is denoted by $\Delta({\cal X})^\circ\equiv\{\bm\lambda=(\lambda_1,\cdots,\lambda_m)\in\Delta({\cal X})\,|\,\lambda_i>0,\,i=1,\cdots,m\}$. Similarly, the set of output probability distributions on the output alphabet ${\cal Y}$ is denoted by $\Delta({\cal Y})\equiv\{Q=(Q_1,\cdots,Q_n)|Q_j\geq0,j=1,\cdots,n,\sum_{j=1}^nQ_j=1\}$.

Let $Q=\bm\lambda\Phi$ be the output distribution for the input distribution $\bm\lambda\in\Delta(\cal X)$, where the representation by components is $Q_j=\sum_{i=1}^m\lambda_iP^i_j,\,j=1,\cdots,n$, then the mutual information is defined by $I(\bm\lambda,\Phi)=\sum_{i=1}^m\sum_{j=1}^n\lambda_iP^i_j\log{P^i_j}/{Q_j}$. The channel capacity $C$ is defined by
\begin{align}
\label{eqn:Cdefinition}
C=\max_{\bm\lambda\in\Delta({\cal X})}I(\bm\lambda,\Phi).
\end{align}
The Kullback-Leibler divergence $D(Q\|Q')$ for two output distributions $Q=(Q_1,\cdots,Q_n),\,Q'=(Q'_1,\cdots,Q'_n)\in\Delta(\cal Y)$ is defined by
\begin{align}
D(Q\|Q')=\sum_{j=1}^nQ_j\log\ds\frac{Q_j}{Q'_j}.
\end{align}
The Kullback-Leibler divergence satisfies $D(Q\|Q')\geq0$, and $D(Q\|Q')=0$ if and only if $Q=Q'$ \cite{csi1}.

An important proposition for investigating the convergence speed of the Arimoto algorithm is the Kuhn-Tucker condition on the input distribution $\bm\lambda=\bm\lambda^\ast$ to achieve the maximum of (\ref{eqn:Cdefinition}).

\medskip

{\it Theorem}\ (Kuhn-Tucker condition) In the maximization problem (\ref{eqn:Cdefinition}), a necessary and sufficient condition for the input distribution $\bm\lambda^\ast=(\lambda^\ast_1,\cdots,\lambda^\ast_m)\in\Delta({\cal X})$ to achieve the maximum is that there is a certain constant $\tilde{C}$ with
\begin{align}
\label{eqn:Kuhn-Tucker}
D(P^i\|\bm\lambda^\ast\Phi)\left\{\begin{array}{ll}=\tilde{C}, & {\mbox{\rm for}}\ i\ {\mbox{\rm with}}\ \lambda^\ast_i>0,\\
\leq \tilde{C}, & {\mbox{\rm for}}\ i\ {\mbox{\rm with}}\ \lambda^\ast_i=0.
\end{array}\right.
\end{align}
In (\ref{eqn:Kuhn-Tucker}), $\tilde{C}$ is equal to the channel capacity $C$.

\medskip

Since this Kuhn-Tucker condition is a necessary and sufficient condition, all the information about the capacity achieving input distribution $\bm\lambda^\ast$ can be derived from this condition.
\section{Arimoto algorithm for calculating channel capacity}
\subsection{Arimoto algorithm\,\cite{ari}}
A sequence of input distributions
\begin{align}
\{\bm\lambda^N=(\lambda^N_1,\cdots,\lambda^N_m)\}_ {N=0,1,\cdots}\subset\Delta({\cal X})
\end{align}
is defined by the Arimoto algorithm as follows. First, let $\bm\lambda^0=(\lambda^0_1,\cdots,\lambda^0_m)$ be an initial distribution taken in $\Delta(\cal X)^\circ$, i.e., $\lambda^0_i>0,\,i=1,\cdots,m$. Then, the Arimoto algorithm is given by the following recurrence formula;
\begin{align}
\lambda^{N+1}_i=\ds\frac{\lambda^N_i\exp D(P^i\|\bm\lambda^N\Phi)}{\ds\sum_{k=1}^m\lambda^N_k\exp D(P^k\|\bm\lambda^N\Phi)},\,i=1,\cdots,m,\,N=0,1,\cdots.\label{eqn:arimotoalgorithm}
\end{align}

On the convergence of this Arimoto algorithm, the following results are obtained in Arimoto\,\cite{ari};

By defining 
\begin{align}
C(N+1,N)\equiv-\ds\sum_{i=1}^m\lambda^{N+1}_i\log\lambda^{N+1}_i+\ds\sum_{i=1}^m\sum_{j=1}^n\lambda^{N+1}_iP^i_j\log\ds\frac{\lambda^N_iP^i_j}{\ds\sum_{k=1}^m\lambda^N_kP^k_j},
\end{align}
they obtained the following theorems;

{\it Theorem A1:} If the initial input distribution $\bm\lambda^0$ is in $\Delta({\cal X})^\circ$, then
\begin{align}
\lim_{N\to\infty}C(N+1,N)=C.
\end{align}

{\it Theorem A2:} If $\bm\lambda^0\in\Delta({\cal X})^\circ$, then
\begin{align}
0\leq C-C(N+1,N)\leq\ds\frac{\log m-h(\bm\lambda^0)}{N},
\end{align}
where $h(\bm\lambda^0)$ is the entropy of $\bm\lambda^0$.

{\it Theorem A3:} If the capacity achieving input distribution $\bm\lambda^\ast$ is in $\Delta({\cal X})^\circ$, then
\begin{align}
0\leq C-C(N+1,N)<K\theta^N,\,N=0,1,\cdots,
\end{align}
where $0\leq\theta<1$ and $K$ is a constant.

In \cite{ari}, they consider the Taylor expansion of $D(\bm\lambda^\ast\|\bm\lambda)$ by $\bm\lambda$, and the Taylor expansion of $D(Q^\ast\|Q)$ by $Q$, however they do not consider the Taylor expansion of the mapping $F:\Delta({\cal X})\to\Delta({\cal X})$, which will be considered in this paper. Further, in the above Theorem A3, they consider only the case $\bm\lambda^\ast\in\Delta({\cal X})^\circ$, where the convergence is exponential.

In Yu\,\cite{yu}, they consider the mapping $F:\Delta({\cal X})\to\Delta({\cal X})$ and the Taylor expansion of $F(\bm\lambda)$ about $\bm\lambda=\bm\lambda^\ast$. They calculate the eigenvalues of the Jacobian matrix $J(\bm\lambda^\ast)$, however they do not consider the Hessian matrix. Further, they consider only the case $\bm\lambda^\ast\in\Delta({\cal X})^\circ$ as in \cite{ari}.

\subsection{Mapping from $\Delta({\cal X})$ to $\Delta({\cal X})$}

Let $F_i(\bm\lambda)$ be the defining function of the Arimoto algorithm (\ref{eqn:arimotoalgorithm}), i.e., 
\begin{align}
F_i(\bm\lambda)=\ds\frac{\lambda_i\exp D(P^i\|\bm\lambda\Phi)}{\ds\sum_{k=1}^m\lambda_k\exp D(P^k\|\bm\lambda\Phi)},\,i=1,\cdots,m.\label{eqn:Arimotofunction}
\end{align}
Define $F(\bm\lambda)=(F_1(\bm\lambda),\cdots,F_m(\bm\lambda))$, then we can consider that $F(\bm\lambda)$ is a differentiable mapping from $\Delta(\cal X)$ to $\Delta(\cal X)$, and (\ref{eqn:arimotoalgorithm}) is represented by
\begin{align}
\bm\lambda^{N+1}=F(\bm\lambda^N).\label{eqn:vectorrecurrence}
\end{align}

In this paper, for the analysis of the convergence speed, we assume 
\begin{align}
{\rm rank}\,\Phi=m.\label{eqn:rankmdefinition}
\end{align}
\begin{lemma}
\label{lem:1}
The capacity achieving input distribution $\bm\lambda^\ast$ is unique.
\end{lemma}
{\bf Proof:} By Csisz\`{a}r\cite{csi1},\,p.137,\,eq.(37), for arbitrary $Q\in\Delta(\cal Y)$,%
\begin{align}
\ds\sum_{i=1}^m\lambda_iD(P^i\|Q)=I(\bm\lambda,\Phi)+D(\bm\lambda\Phi\|Q).\label{eqn:CKequality}
\end{align}
By the assumption (\ref{eqn:rankmdefinition}), we see that there exists $Q^0\in\Delta(\cal Y)$ \cite{nak2} with 

\begin{align}
D(P^1\|Q^0)=\cdots=D(P^m\|Q^0)\equiv C^0.
\end{align}
Substituting $Q=Q^0$ into (\ref{eqn:CKequality}), we have $C^0=I(\bm\lambda,\Phi)+D(\bm\lambda\Phi\|Q^0)$. Because $C^0$ is a constant,
\begin{align}
\max_{\bm\lambda\in\Delta(\cal X)}I(\bm\lambda,\Phi)\Longleftrightarrow\min_{\bm\lambda\in\Delta(\cal X)}D(\bm\lambda\Phi\|Q^0).\label{eqn:maxequalmin}
\end{align}
Define $V\equiv\{\bm\lambda\Phi\,|\,\bm\lambda\in\Delta(\cal X)\}$, then $V$ is a closed convex set, thus by Cover\,\cite{cov},\,p.297,\,Theorem 12.6.1, $Q=Q^\ast$ that achieves $\min_{Q\in V}D(Q\|Q^0)$ exists and is unique. By the assumption (\ref{eqn:rankmdefinition}), the mapping $\Delta\ni\bm\lambda\mapsto\bm\lambda\Phi\in V$ is one to one, therefore, $\bm\lambda^\ast$ with $Q^\ast=\bm\lambda^\ast\Phi$ is unique.\hfill$\blacksquare$
\begin{remark}
\rm Due to the equivalence (\ref{eqn:maxequalmin}), the Arimoto algorithm can be obtained by Csisz\`{a}r \cite{csi2}, Chapter 4, ``Minimizing information distance from a single measure'', Theorem 5.
\end{remark}
\begin{lemma}
\label{lem:2}
The capacity achieving input distribution $\bm\lambda^\ast$ is the fixed point of the mapping $F(\bm\lambda)$ in $(\ref{eqn:vectorrecurrence})$. That is, $\bm\lambda^\ast=F(\bm\lambda^\ast)$.
\end{lemma}
{\bf Proof:} In the Kuhn-Tucker condition (\ref{eqn:Kuhn-Tucker}), let us define $m_1$ as the number of indices $i$ with $\lambda^\ast_i>0$, i.e., 
\begin{align}
\lambda^\ast_i\left\{\begin{array}{ll}>0, & i=1,\cdots,m_1,\\=0, & i=m_1+1,\cdots,m,\end{array}\right.\label{eqn:m1definition}
\end{align}
then 
\begin{align}
D(P^i\|\bm\lambda^\ast\Phi)\left\{\begin{array}{ll}=C, & i=1,\cdots,m_1,\\\leq C, & i=m_1+1,\cdots,m.\end{array}\right.
\end{align}
We have
\begin{align}
\ds\sum_{k=1}^m\lambda^\ast_k\exp D(P^k\|\bm\lambda^\ast\Phi)=\ds\sum_{k=1}^{m_1}\lambda^\ast_ke^C=e^C,\label{eqn:yobitekikeisan}
\end{align}
hence by (\ref{eqn:Arimotofunction}),\,(\ref{eqn:m1definition}),\,(\ref{eqn:yobitekikeisan}),
\begin{align}
F_i(\bm\lambda^\ast)&=\left\{\begin{array}{ll}e^{-C}\lambda^\ast_ie^C, & i=1,\cdots,m_1,\\0, & i=m_1+1,\cdots,m,\end{array}\right.\\
&=\lambda^\ast_i,\,i=1,\cdots,m,
\end{align}
which shows $F(\bm\lambda^\ast)=\bm\lambda^\ast$.\hfill$\blacksquare$

\medskip

The sequence $\bm\lambda^N$ of the Arimoto algorithm converges to the fixed point $\bm\lambda^\ast$, i.e., 
\begin{align}
\bm\lambda^N\to\bm\lambda^\ast,\,N\to\infty.\label{eqn:lambdaNconverges}
\end{align}
We will investigate the convergence speed by using the Taylor expansion of $F(\bm\lambda)$ about $\bm\lambda=\bm\lambda^\ast$.
\subsection{Type of index}
Now, we classify the indices $i\,(i=1,\cdots,m)$ in the Kuhn-Tucker condition (\ref{eqn:Kuhn-Tucker}) in more detail into the following 3 types;
\begin{align}
\label{eqn:Kuhn-Tucker2}
D(P^i\|\bm\lambda^\ast\Phi)\left\{\begin{array}{ll}=C, & {\mbox{\rm for}}\ i\ {\mbox{\rm with}}\ \lambda^\ast_i>0\ {\rm (type\ I)},\\
=C, & {\mbox{\rm for}}\ i\ {\mbox{\rm with}}\ \lambda^\ast_i=0\ {\rm (type\ II)},\\
<C, & {\mbox{\rm for}}\ i\ {\mbox{\rm with}}\ \lambda^\ast_i=0\ {\rm (type\ III)}.
\end{array}\right.
\end{align}
Let us define the sets of indices as follows;
\begin{align}
&{\rm all\ the\ indices:}\ {\cal I}\equiv\{1,\cdots,m\},\label{eqn:allset}\\
&{\rm type\ I\ indices:}\ {\cal I}_{\rm I}\equiv\{1,\cdots,m_1\},\label{eqn:type1set}\\
&{\rm type\ II\ indices:}\ {\cal I}_{\rm II}\equiv\{m_1+1,\cdots,m_1+m_2\},\label{eqn:type2set}\\
&{\rm type\ III\ indices:}\ {\cal I}_{\rm III}\equiv\{m_1+m_2+1,\cdots,m\}.\label{eqn:type3set}
\end{align}
$|{\cal I}|=m$, $|{\cal I}_{\rm I}|=m_1$, $|{\cal I}_{\rm II}|=m_2$, $|{\cal I}_{\rm III}|=m-m_1-m_2\equiv m_3$. We have ${\cal I}={\cal I}_{\rm I}\cup{\cal I}_{\rm II}\cup{\cal I}_{\rm III}$ and $m=m_1+m_2+m_3$.

${\cal I}_{\rm I}$ is not empty and $|{\cal I}_{\rm I}|=m_1\geq2$ for any channel matrix, but ${\cal I}_{\rm II}$ and ${\cal I}_{\rm III}$ may be empty for some channel matrix.
\subsection{Examples of convergence speed}
Let us consider the difference of convergence speed  of the Arimoto algorithm depending on the channel matrices.

For many channel matrices $\Phi$, the convergence is exponential, but for some special $\Phi$ the convergence is very slow. Let us consider the following examples taking types I, II, III into account, where the input alphabet size $m=3$ and the output alphabet size $n=3$.

\begin{example}
\label{exa:1}
\rm (only type I) If only type I indices exist, then $\lambda^\ast_i>0,\,i=1,2,3$, hence $Q^\ast\equiv\bm\lambda^\ast\Phi$ is in the interior of $\triangle P^1P^2P^3$. As a concrete channel matrix of this example, let us consider
\begin{align}
\label{eqn:Phi1}
\Phi^{(1)}=\begin{pmatrix}
\,0.800 & 0.100 & 0.100\,\\
\,0.100 & 0.800 & 0.100\,\\
\,0.250 & 0.250 & 0.500\,
\end{pmatrix}.
\end{align}
For this $\Phi^{(1)}$, we have $\bm\lambda^\ast=(0.431,0.431,0.138)$ and $Q^\ast=(0.422,0.422,0.156)$. See Fig.\ref{fig:1}. The vertices of the large triangle in Fig.\ref{fig:1} are the output probability distributions $\bm{e}^1=(1,0,0),\,\bm{e}^2=(0,1,0),\,\bm{e}^3=(0,0,1)$. We have $D(P^i\|Q^\ast)=C,\,i=1,2,3$, then considering the analogy to Euclidean geometry, $\triangle P^1P^2P^3$ can be regarded as an ``acute triangle''.
\begin{figure}[t]
\begin{center}
\begin{overpic}[width=8.8cm]{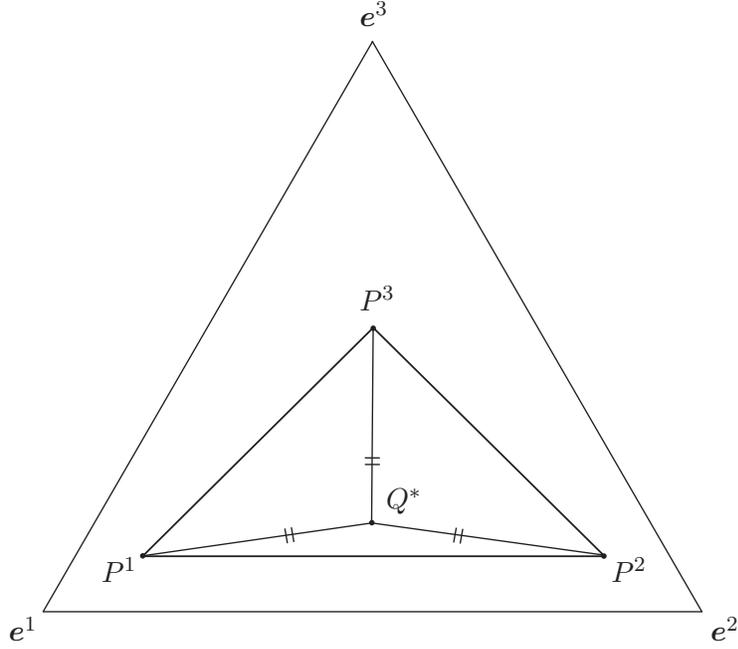}
%\begin{overpic}[width=8cm,grid]{./figure/triangle1.eps}
\put(-5,-4){$\bm{e}^1$}
\put(101,-4){$\bm{e}^2$}
\put(48,89){$\bm{e}^3$}
\put(52,16){$Q^\ast$}
\put(9,5){$P^1$}
\put(86,5){$P^2$}
\put(48,46){$P^3$}
\end{overpic}
\medskip
\caption{Positional relation of row vectors $P^1,P^2,P^3$ of $\Phi^{(1)}$ and $Q^\ast$ in Example \ref{exa:1}}
\label{fig:1}
\end{center}
\end{figure}
\end{example}

\begin{example}
\label{exa:2}
\rm (types I and II) If there are type I and type II indices, we can assume $\lambda^\ast_1>0,\lambda^\ast_2>0,\lambda^\ast_3=0$ without loss of generality, hence $Q^\ast$ is on the side $P^1P^2$ and $D(P^i\|Q^\ast)=C,\,i=1,2,3$. As a concrete channel matrix of this example, let us consider
\begin{align}
\label{eqn:Phi2}
\Phi^{(2)}=\begin{pmatrix}
\,0.800 & 0.100 & 0.100\,\\
\,0.100 & 0.800 & 0.100\,\\
\,0.300 & 0.300 & 0.400\,
\end{pmatrix}.
\end{align}
For this $\Phi^{(2)}$, we have $\bm\lambda^\ast=(0.500,0.500,0.000)$ and $Q^\ast=(0.450,0.450,0.100)$. See Fig.\ref{fig:2}. Considering the analogy to Euclidean geometry, $\triangle P^1P^2P^3$ can be regarded as a ``right triangle''. 

\begin{figure}[t]
\begin{center}
 \medskip
\begin{overpic}[width=8.8cm]{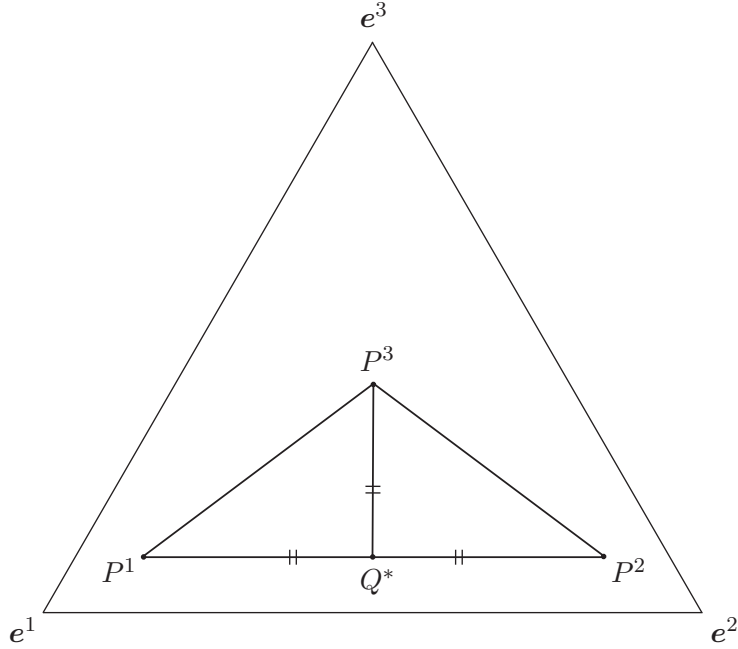}
%\begin{overpic}[width=8cm,grid]{./figure/triangle2.eps}
\put(-5,-4){$\bm{e}^1$}
\put(101,-4){$\bm{e}^2$}
\put(48,89){$\bm{e}^3$}
\put(48,4){$Q^\ast$}
\put(9,5){$P^1$}
\put(86,5){$P^2$}
\put(48,37){$P^3$}
\end{overpic}
\medskip
\caption{Positional relation of row vectors $P^1,P^2,P^3$ of $\Phi^{(2)}$ and $Q^\ast$ in Example \ref{exa:2}}
\label{fig:2}
\end{center}
\end{figure}
\end{example}

\begin{example}
\label{exa:3}
\rm (types I and III) If there are type I and type III indices, we can assume $\lambda^\ast_1>0,\lambda^\ast_2>0,\lambda^\ast_3=0$ without loss of generality, hence $Q^\ast$ is on the side $P^1P^2$ and $C=D(P^1\|Q^\ast)=D(P^2\|Q^\ast)>D(P^3\|Q^\ast)$. As a concrete channel matrix of this example, let us consider
\begin{align}
\label{eqn:Phi3}
\Phi^{(3)}=\begin{pmatrix}
\,0.800 & 0.100 & 0.100\,\\
\,0.100 & 0.800 & 0.100\,\\
\,0.350 & 0.350 & 0.300\,
\end{pmatrix}.
\end{align}
For this $\Phi^{(3)}$, we have $\bm\lambda^\ast=(0.500,0.500,0.000)$ and $Q^\ast=(0.450,0.450,0.100)$. See Fig.\ref{fig:3}. Considering the analogy to Euclidean geometry, $\triangle P^1P^2P^3$ can be regarded as an ``obtuse triangle''. 
\begin{figure}[t]
\begin{center}
\begin{overpic}[width=8.8cm]{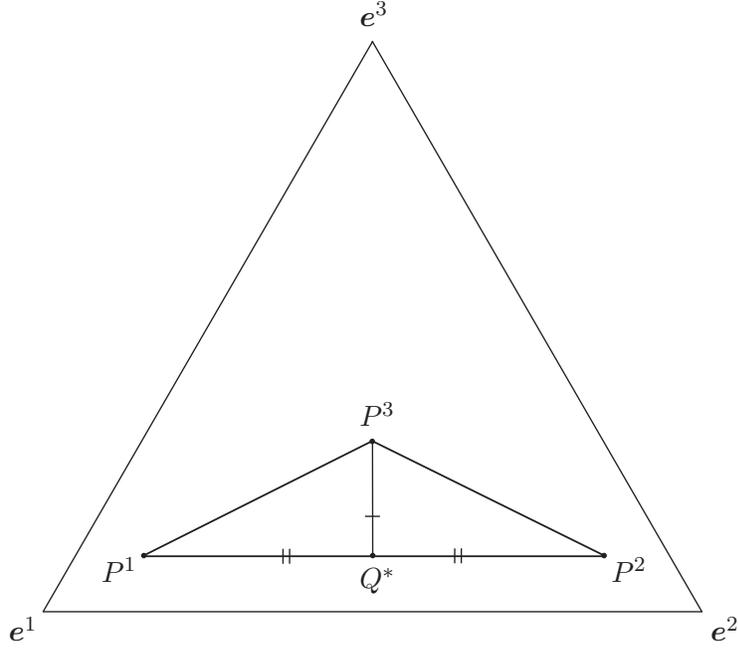}
%\begin{overpic}[width=8cm,grid]{./figure/triangle3.eps}
\put(-5,-4){$\bm{e}^1$}
\put(101,-4){$\bm{e}^2$}
\put(48,89){$\bm{e}^3$}
\put(48,4){$Q^\ast$}
\put(9,5){$P^1$}
\put(86,5){$P^2$}
\put(48,28.5){$P^3$}
\end{overpic}
\medskip
\caption{Positional relation of row vectors $P^1,P^2,P^3$ of $\Phi^{(3)}$ and $Q^\ast$ in Example \ref{exa:3}}
\label{fig:3}
\end{center}
\end{figure}
\end{example}

\begin{figure}[t]
\begin{center}
\begin{overpic}[width=9cm]{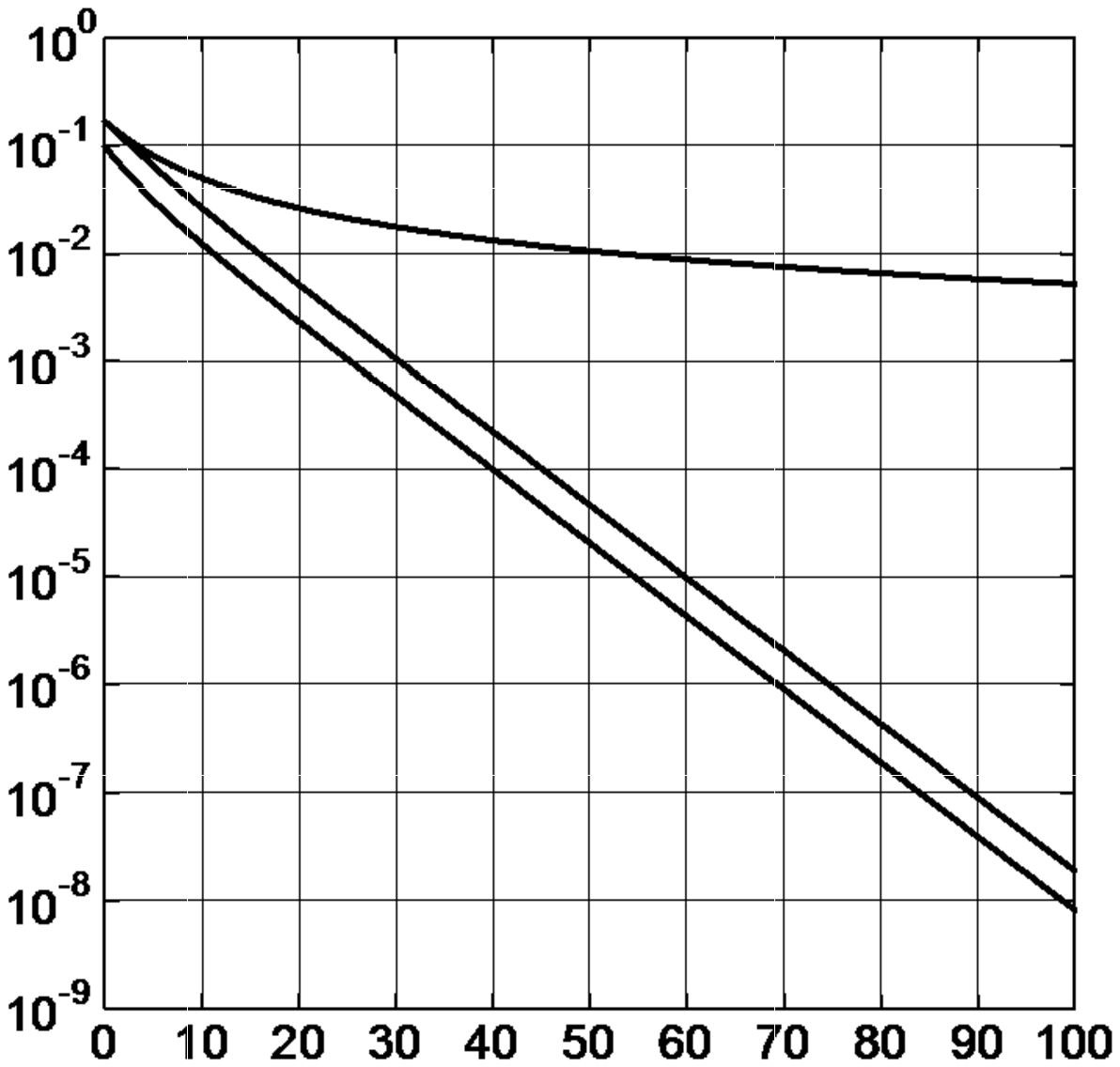}
%\begin{overpic}[width=8.2cm,grid]{./figure/lambda1.eps}
%
\put(57,74){\rotatebox{60}{$\leftarrow$}}
\put(62,78){Example 2}
\put(62,74){(types I and II)}
\put(58,47.5){\rotatebox{60}{$\leftarrow$}}
\put(62,50){Example 3}
\put(62,46){(types I and III)}
\put(52,40){\rotatebox{60}{$\rightarrow$}}
\put(39,36){Example 1}
\put(39,31.5){(only type I)}
\put(54,-3){$N$}
\put(-6,45){\rotatebox{90}{$|\lambda^N_1-\lambda^\ast_1|$}}
\end{overpic}
\caption{Comparison of the convergence speed in Examples \ref{exa:1},\ref{exa:2},\ref{exa:3}}
\label{fig:4}
\end{center}
\end{figure}

For the above $\Phi^{(1)},\Phi^{(2)},\Phi^{(3)}$, Fig.\ref{fig:4} shows the state of convergence of $|\lambda^N_1-\lambda^\ast_1|\to0$. By this Figure, we see that in Examples 1 and 3 the convergence is exponential, while in Example 2 the convergence is slower than exponential.

From the above three examples, it is inferred that the Arimoto algorithm converges very slowly when type II index exists, and converges exponentially when type II index does not exist. We will analyze this phenomenon in the following.
\section{Taylor expansion of $F(\bm\lambda)$ about $\bm\lambda=\bm\lambda^\ast$}
We will examine the convergence speed of the Arimoto algorithm by the Taylor expansion of $F(\bm\lambda)$ about the fixed point $\bm\lambda=\bm\lambda^\ast$. Taylor expansion of the function $F(\bm\lambda)=(F_1(\bm\lambda),\cdots,F_m(\bm\lambda))$ about $\bm\lambda=\bm\lambda^\ast$ is
\begin{align}
F(\bm\lambda)=F(\bm\lambda^\ast)+(\bm\lambda-\bm\lambda^\ast)J(\bm\lambda^\ast)+\ds\frac{1}{2!}(\bm\lambda-\bm\lambda^\ast)H(\bm\lambda^\ast)\,^t(\bm\lambda-\bm\lambda^\ast)+o(\|\bm\lambda-\bm\lambda^\ast\|^2),\label{eqn:Taylortenkai1}
\end{align}
where ${^t}\bm\lambda$ denotes the transpose of $\bm\lambda$ and $\|\bm\lambda\|$ denotes the Euclidean norm $\|\bm\lambda\|=\left(\lambda_1^2+\cdots+\lambda_m^2\right)^{1/2}$.

In (\ref{eqn:Taylortenkai1}), $J(\bm\lambda^\ast)$ is the Jacobian matrix at $\bm\lambda=\bm\lambda^\ast$, i.e.,
\begin{align}
J(\bm\lambda^\ast)&=\left(\left.\ds\frac{\partial F_i}{\partial\lambda_{i'}}\right|_{\bm\lambda=\bm\lambda^\ast}\right)_{i',i=1,\cdots,m}.\label{eqn:Jacobiseibun}
\end{align}

We consider in this paper that the input probability distribution $\bm\lambda$ is a row vector, thus the Jacobian matrix $J(\bm\lambda^\ast)$ is such as
\begin{align}
&\hspace{30mm}\leftarrow i\rightarrow\nonumber\\[0mm]
J(\bm\lambda^\ast)&=\begin{array}{c}\uparrow\\ i'\\\downarrow\end{array}
\hspace{-1mm}\begin{pmatrix}
\,\left.\ds\frac{\partial F_1}{\partial\lambda_1}\right|_{\bm\lambda=\bm\lambda^\ast} & \cdots & \left.\ds\frac{\partial F_m}{\partial\lambda_1}\right|_{\bm\lambda=\bm\lambda^\ast}\,\\
\vdots & & \vdots\\
\,\left.\ds\frac{\partial F_1}{\partial\lambda_m}\right|_{\bm\lambda=\bm\lambda^\ast} & \cdots & \left.\ds\frac{\partial F_m}{\partial\lambda_m}\right|_{\bm\lambda=\bm\lambda^\ast}\,
\end{pmatrix}\in\mathbb R^{m\times m},\label{eqn:rowvectorJacobimatrix}
\end{align}
i.e., $\partial F_i/\partial\lambda_{i'}|_{\bm\lambda=\bm\lambda^\ast}$ is the $(i',i)$ component. Note that our $J(\bm\lambda^\ast)$ is the transpose of a usual Jacobian matrix corresponding to column vector. 

Because $\sum_{i=1}^mF_i(\bm\lambda)=1$ by (\ref{eqn:Arimotofunction}), we have by (\ref{eqn:rowvectorJacobimatrix}),
\begin{lemma}
\label{lem:rowsumofJis0}
Every row sum of $J(\bm\lambda^\ast)$ is equal to $0$.
\end{lemma}

\medskip

In (\ref{eqn:Taylortenkai1}), $H(\bm\lambda^\ast)\equiv(H_1(\bm\lambda^\ast),\cdots,H_m(\bm\lambda^\ast))$, where $H_i(\bm\lambda^\ast)$ is the Hessian matrix of $F_i$ at $\bm\lambda=\bm\lambda^\ast$, i.e.,
\begin{align}
H_i(\bm\lambda^\ast)=\left(\left.\ds\frac{\partial^2F_i}{\partial\lambda_{i'}\partial\lambda_{i''}}\right|_{\bm\lambda=\bm\lambda^\ast}\right)_{i',i''=1,\cdots,m},\label{eqn:Hesseseibun}
\end{align}
and $(\bm\lambda-\bm\lambda^\ast)H(\bm\lambda^\ast)\,^t(\bm\lambda-\bm\lambda^\ast)$ is an abbreviated expression of the $m$ dimensional vector $((\bm\lambda-\bm\lambda^\ast)H_1(\bm\lambda^\ast)\,^t(\bm\lambda-\bm\lambda^\ast),\cdots,(\bm\lambda-\bm\lambda^\ast)H_m(\bm\lambda^\ast)\,^t(\bm\lambda-\bm\lambda^\ast)).$

\begin{remark}
\label{rem:justify}
\rm $\lambda_1,\cdots,\lambda_m$ satisfy the constraint $\sum_{i=1}^m\lambda_i=1$, but in (\ref{eqn:Taylortenkai1}),\,(\ref{eqn:Jacobiseibun}),\,(\ref{eqn:Hesseseibun}) we consider $\lambda_1,\cdots,\lambda_m$ as independent variables to have the Taylor series approximation (\ref{eqn:Taylortenkai1}). This approximation is justified as follows. By the Kuhn-Tucker condition (\ref{eqn:Kuhn-Tucker}), $D(P^i\|Q^\ast)\leq C<\infty,\,i=1,\cdots,m$, hence by the assumption put below (\ref{eqn:thechannelmatrix}), we have $Q^\ast_j>0,\,j=1,\cdots,n$. See \cite{ari}. For $\epsilon>0$, define ${\cal Q}^\ast_\epsilon\equiv\{Q=(Q_1,\cdots,Q_n)\in{\mathbb R}^n\,|\,\|Q-Q^\ast\|<\epsilon\}$, i.e., ${\cal Q}^\ast_\epsilon$ is an open ball in $\mathbb{R}^n$ centered at $Q^\ast$ with radius $\epsilon$. Note that $Q\in{\cal Q}^\ast_\epsilon$ is free from the constraint $\sum_{j=1}^nQ_j=1$. Taking $\epsilon>0$ sufficiently small, we can have $Q_j>0,j=1,\cdots,n$, for any $Q\in{\cal Q}^\ast_\epsilon$. The function $F(\bm\lambda)$ is defined for $\bm\lambda$ with $\left(\bm\lambda\Phi\right)_j>0,\,j=1,\cdots,n$, even if some $\lambda_i<0$. Therefore, the domain of definition of $F(\bm\lambda)$ can be extended to $\Phi^{-1}\left({\cal Q}^\ast_\epsilon\right)\subset\mathbb{R}^m$, where $\Phi^{-1}\left({\cal Q}^\ast_\epsilon\right)$ is the inverse image of ${\cal Q}^\ast_\epsilon$ by the mapping $\mathbb{R}^m\ni\bm\lambda\to\bm\lambda\Phi\in\mathbb{R}^n$. $\Phi^{-1}\left({\cal Q}^\ast_\epsilon\right)$ is an open neighborhood of $\bm\lambda^\ast$ in $\mathbb{R}^m$. Then $F(\bm\lambda)$ is a function of $\bm\lambda=(\lambda_1,\cdots,\lambda_m)\in\Phi^{-1}\left({\cal Q}^\ast_\epsilon\right)$ as independent variables (free from the constraint $\sum_{i=1}^m\lambda_i=1$). We can consider (\ref{eqn:Taylortenkai1}) to be the Taylor expansion by independent variables $\lambda_1,\cdots,\lambda_m$, then substituting $\bm\lambda\in\Delta({\cal X})\cap\Phi^{-1}\left({\cal Q}^\ast_\epsilon\right)$ into (\ref{eqn:Taylortenkai1}) to obtain the approximation for $F(\bm\lambda)$ about $\bm\lambda=\bm\lambda^\ast$.
\end{remark}

\medskip

Now, substituting $\bm\lambda=\bm\lambda^N$ into (\ref{eqn:Taylortenkai1}), then by $F(\bm\lambda^\ast)=\bm\lambda^\ast$ and $F(\bm\lambda^N)=\bm\lambda^{N+1}$, we have
\begin{align}
\bm\lambda^{N+1}=\bm\lambda^\ast+(\bm\lambda^N-\bm\lambda^\ast)J(\bm\lambda^\ast)+\ds\frac{1}{2!}(\bm\lambda^N-\bm\lambda^\ast)H(\bm\lambda^\ast)\,^t(\bm\lambda^N-\bm\lambda^\ast)+o(\|\bm\lambda^N-\bm\lambda^\ast\|^2).\label{eqn:Taylortenkai2}
\end{align}
Then, by putting $\bm\mu^N\equiv\bm\lambda^N-\bm\lambda^\ast$, (\ref{eqn:Taylortenkai2}) becomes
\begin{align}
\bm\mu^{N+1}=\bm\mu^NJ(\bm\lambda^\ast)+\ds\frac{1}{2!}\bm\mu^NH(\bm\lambda^\ast)\,{^t}\bm\mu^N+o\left(\|\bm\mu^N\|^2\right).\label{eqn:Taylortenkai3}
\end{align}

By (\ref{eqn:lambdaNconverges}), we will investigate the convergence
\begin{align}
\bm\mu^N\to\bm0,\,N\to\infty,
\end{align}
based on the Taylor expansion (\ref{eqn:Taylortenkai3}). Let
\begin{align}
\mu^N_i\equiv\lambda^N_i-\lambda^\ast_i,\,i=1,\cdots,m,\label{eqn:muislambdaminuslambda}
\end{align}
denote the components of $\bm\mu^N=\bm\lambda^N-\bm\lambda^\ast$, and write $\bm\mu^N$ by components as $\bm\mu^N=(\mu^N_1,\cdots,\mu^N_m)$, then we have 
\begin{align}
\sum_{i=1}^m\mu^N_i=0,\,N=0,1,\cdots,\label{eqn:musumiszero}
\end{align}
because $\sum_{i=1}^m\lambda^N_i=\sum_{i=1}^m\lambda^\ast_i=1$.
\subsection{Basic analysis for fast and slow convergence}
For the investigation of the convergence speed, we consider the following simple case.

Let us define a real sequence $\{\mu^N\}_{N=0,1,\cdots}\subset{\mathbb R}$ by the recurrence formula;
\begin{align}
\mu^{N+1}&=\theta\mu^N-\rho\left(\mu^N\right)^2,\,N=0,1,\cdots,\label{eqn:sequenceaN}\\
0&<\theta\leq1,\,\rho>0,\,0<\mu^0<\theta/\rho.
\end{align}

If $0<\theta<1$, then we have $0<\mu^{N+1}<\theta\mu^N<\cdots<\theta^{N+1}\mu^0$, hence $\mu^N$ decays exponentially.

\begin{figure}[t]
\begin{center}
\begin{overpic}[width=8cm]{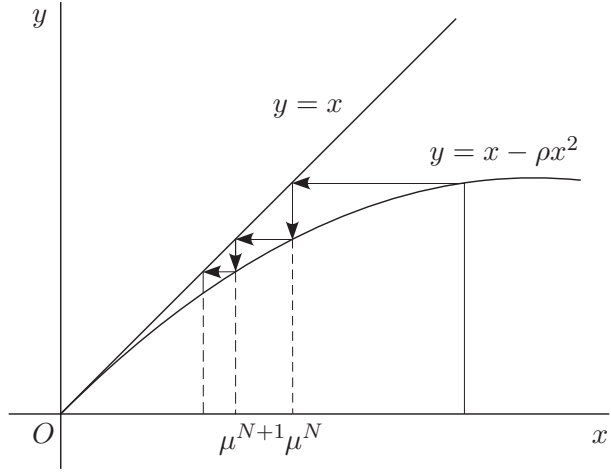}
%\begin{overpic}[width=9cm,grid]{./figure/slow_decay.eps}
\put(46,4){$\mu^N$}
\put(35,4){$\mu^{N+1}$}
\put(97,5){$x$}
\put(4,75){$y$}
\put(44,60){$y=x$}
\put(70,52){$y=x-\rho x^2$}
\put(4,5){$O$}
\end{overpic}
\medskip
\caption{Convergence of the sequence defined by $\mu^{N+1}=\mu^N-\rho\left(\mu^N\right)^2$}
\label{fig:5}
\end{center}
\end{figure}

While, if $\theta=1$, (\ref{eqn:sequenceaN}) becomes $\mu^{N+1}=\mu^N-\rho\left(\mu^N\right)^2,\,\rho>0$. This recurrence formula cannot be solved explicitly, however, we see the state of convergence by Fig.\ref{fig:5}. 
Because the differential coefficient of the function $y=x-\rho x^2$ at $x=0$ is 1, the convergence speed is very slow. In fact, this convergence is slower than exponential. From Lemma \ref{lem:6} in section \ref{sec:m3narbitray} below, we will see that the convergence speed is the $1/N$ order and $\lim_{N\to\infty}N\mu^N=1/\rho$.

\subsection{On Jacobian matrix $J(\bm\lambda^\ast)$}
Let us consider the Jacobian matrix $J(\bm\lambda^\ast)$ for any $m,n$. We are assuming ${\rm rank}\,\Phi=m$ in (\ref{eqn:rankmdefinition}), hence $m\leq n$. 

We will calculate the components (\ref{eqn:Jacobiseibun}) of $J(\bm\lambda^\ast)$.

Defining
\begin{align}
&D_i\equiv D(P^i\|\bm\lambda\Phi),\,i=1,\cdots,m,\\
&F_i\equiv F_i(\bm\lambda),\,i=1,\cdots,m,
\end{align}
we can write (\ref{eqn:Arimotofunction}) as
\begin{align}
F_i=\ds\frac{\lambda_ie^{D_i}}{\ds\sum_{k=1}^m\lambda_ke^{D_k}},\,i=1,\cdots,m.\label{eqn:teigikansuFi}
\end{align}
From (\ref{eqn:teigikansuFi}),
\begin{align}
F_i\ds\sum_{k=1}^m\lambda_ke^{D_k}=\lambda_ie^{D_i},\label{eqn:Fibunboharau}
\end{align}
then differentiating the both sides of (\ref{eqn:Fibunboharau}) by $\lambda_{i'}$, we have
\begin{align}
\ds\frac{\partial F_i}{\partial\lambda_{i'}}\ds\sum_{k=1}^m\lambda_ke^{D_k}+F_i\ds\frac{\partial}{\partial\lambda_{i'}}\ds\sum_{k=1}^m\lambda_ke^{D_k}=\delta_{i'i}e^{D_i}+\lambda_ie^{D_i}\ds\frac{\partial D_i}{\partial\lambda_{i'}},\label{eqn:dFi}
\end{align}
where $\delta_{i'i}$ is the Kronecker delta.

Before substituting $\bm\lambda=\bm\lambda^\ast=(\lambda^\ast_1,\cdots,\lambda^\ast_m)$ into the both sides of (\ref{eqn:dFi}), we define the following symbols. Remember that the integer $m_1$ was defined in (\ref{eqn:m1definition}). See also (\ref{eqn:type1set}).

Let us define
\begin{align}
Q^\ast&\equiv Q(\bm\lambda^\ast)=\bm\lambda^\ast\Phi,\\
Q_j^\ast&\equiv Q(\bm\lambda^\ast)_j=\ds\sum_{i=1}^m\lambda_i^\ast P_j^i=\ds\sum_{i=1}^{m_1}\lambda_i^\ast P_j^i,\,j=1,\cdots,n,\\
D_i^\ast&\equiv D(P^i\|Q^\ast),\,i=1,\cdots,m,\\
D_{i',i}^\ast&\left.\equiv\ds\frac{\partial D_i}{\partial\lambda_{i'}}\right|_{\bm\lambda=\bm\lambda^\ast},\,i',i=1,\cdots,m,\label{eqn:Diidefinition}\\
F_i^\ast&\equiv F_i(\bm\lambda^\ast),\,i=1,\cdots,m.
\end{align}
\begin{lemma}
\label{lem:3}
\begin{align}
&\left.\ds\sum_{k=1}^m\lambda_ke^{D_k}\right|_{\bm\lambda=\bm\lambda^\ast}=e^C,\label{eqn:lem3-1}\\[2mm]
&\ds\frac{\partial D_i}{\partial\lambda_{i'}}=-\ds\sum_{j=1}^n\ds\frac{P_j^{i'}P_j^i}{Q_j},\,i',i=1,\cdots,m,\label{eqn:lem3-2}\\[2mm]
&\left.\ds\frac{\partial}{\partial\lambda_{i'}}\ds\sum_{k=1}^m\lambda_ke^{D_k}\right|_{\bm\lambda=\bm\lambda^\ast}=e^{D_{i'}^\ast}-e^C,\,i'=1,\cdots,m,\label{eqn:lem3-3}\\[2mm]
&F^\ast_i=\lambda^\ast_i,\,i=1,\cdots,m.\label{eqn:lem3-4}
\end{align}
\end{lemma}
{\bf Proof:} 
We have (\ref{eqn:lem3-1}),\,(\ref{eqn:lem3-2}) by simple calculation. See (\ref{eqn:yobitekikeisan}). (\ref{eqn:lem3-4}) is the result of Lemma \ref{lem:2}. (\ref{eqn:lem3-3}) is proved as follows;
\begin{align*}
\left.\ds\frac{\partial}{\partial\lambda_{i'}}\ds\sum_{k=1}^m\lambda_ke^{D_k}\right|_{\bm\lambda=\bm\lambda^\ast}
&=\left.\ds\sum_{k=1}^m\left(\delta_{i'k}e^{D_k}+\lambda_ke^{D_k}\ds\frac{\partial D_k}{\partial\lambda_{i'}}\right)\right|_{\bm\lambda=\bm\lambda^\ast}\\
&=e^{D_{i'}^\ast}+\ds\sum_{k=1}^{m_1}\lambda_k^\ast e^C\left(-\ds\sum_{j=1}^n\ds\frac{P_j^kP_j^{i'}}{Q_j^\ast}\right)\\
&=e^{D_{i'}^\ast}-e^C\ds\sum_{j=1}^nP_j^{i'}\ds\frac{1}{Q_j^\ast}\ds\sum_{k=1}^{m_1}\lambda_k^\ast P_j^k\\
&=e^{D_{i'}^\ast}-e^C.
\end{align*}
Note that $Q^\ast_j>0,\,j=1,\cdots,n$, from Remark \ref{rem:justify}.\hfill$\blacksquare$

\medskip

Substituting the results of Lemma \ref{lem:3} into (\ref{eqn:dFi}), we have
\begin{align}
\left.\ds\frac{\partial F_i}{\partial\lambda_{i'}}\right|_{\bm\lambda=\bm\lambda^\ast}e^C+\lambda_i^\ast\left(e^{D_{i'}^\ast}-e^C\right)=\delta_{i'i}e^{D^\ast_i}+\lambda_i^\ast e^{D^\ast_i}D^\ast_{i,i'}.
\end{align}

Consequently, we have
\begin{theorem}
\label{the:1}
\begin{align}
\left.\ds\frac{\partial F_i}{\partial\lambda_{i'}}\right|_{\bm\lambda=\bm\lambda^\ast}&=e^{D_i^\ast-C}\left(\delta_{i'i}+\lambda_i^\ast D^\ast_{i',i}\right)+\lambda^\ast_i\left(1-e^{D^\ast_{i'}-C}\right),\,i',i\in{\cal I},\nonumber\\[1mm]
&=\left\{\begin{array}{l}
\delta_{i'i}+\lambda_i^\ast\left(D^\ast_{i',i}+1-e^{D_{i'}^\ast-C}\right),\,i'\in{\cal I},\,i\in{\cal I}_{\rm I},\\[2mm]
\delta_{i'i},\,i'\in{\cal I},\,i\in{\cal I}_{\rm II},\\[2mm]
e^{D^\ast_i-C}\delta_{i'i},\,i'\in{\cal I},\,i\in{\cal I}_{\rm III},
\end{array}\right.\label{eqn:theorem1-1}
\end{align}
where the sets of indices ${\cal I}$, ${\cal I}_{\rm I}$, ${\cal I}_{\rm II}$, ${\cal I}_{\rm III}$ were defined in $(\ref{eqn:allset})$-$(\ref{eqn:type3set})$. Note that $D^\ast_i=C$ for $i\in{\cal I}_{\rm I}\cup{\cal I}_{\rm II}$ and $\lambda^\ast_i=0$ for $i\in{\cal I}_{\rm II}\cup{\cal I}_{\rm III}$.
\end{theorem}
%
%
%Especially, if $D_i^\ast=C$ for all $1\leq i\leq m$, then
%
%\begin{align}
%\hspace{-1mm}\left.\ds\frac{\partial F_i}{\partial\lambda_{i'}}\right|_{\bm\lambda=\bm\lambda^\ast}\!\!\!=\left\{\begin{array}{l}
%\delta_{i'i}+\lambda_i^\ast D_{i',i}^\ast,\\
%\ \ i'=1,\cdots,m,\,i=1,\cdots,m_1,\\
%\delta_{i'i}, \\
%\ \ i'=1,\cdots,m,\,i=m_1+1,\cdots,m.
%\end{array}\right.\label{eqn:theorem1-2}
%\end{align}
%\end{theorem}
%
\subsection{Eigenvalues of Jacobian matrix $J(\bm\lambda^\ast)$}
From (\ref{eqn:theorem1-1}), we see that the Jacobian matrix $J(\bm\lambda^\ast)$ is of the form
\begin{align}
&J(\bm\lambda^\ast)\equiv\begin{pmatrix}
\,J^{\rm I} & O & O\,\\[1mm]
\,\ast & J^{\rm II} & O\,\\[1mm]
\,\ast & O & J^{\rm III}
\end{pmatrix},\label{eqn:J1AJ2}\\
&J^{\rm I}\in{\mathbb R}^{m_1\times m_1},\label{eqn:Jstructure1}\\
&J^{\rm II}=I\,({\rm the\ identity\ matrix})\in{\mathbb R}^{m_2\times m_2},\label{eqn:Jstructure2}\\
&J^{\rm III}={\rm diag}\left(e^{D^\ast_{m_1+m_2+1}-C},\cdots,e^{D^\ast_m-C}\right)\in{\mathbb R}^{m_3\times m_3},\label{eqn:Jstructure3}\\
&\text{\rm where}\ D^\ast_{m_1+m_2+1}<C,\cdots,D^\ast_m<C\ \text{\rm by\ type III\ in}\ (\ref{eqn:Kuhn-Tucker2}),\nonumber\\
&O\ {\rm denotes\ the\ zero\ matrix\ of\ appropriate\ size.}\nonumber
\end{align}

Let $\{\theta_1,\cdots,\theta_m\}\equiv\{\theta_i\,|\,i\in{\cal I}\}$ be the set of eigenvalues of $J(\bm\lambda^\ast)$. By (\ref{eqn:J1AJ2}), the eigenvalues of $J(\bm\lambda^\ast)$ are the eigenvalues of $J^{\rm I}$, $J^{\rm II}$, $J^{\rm III}$, hence we can put

$\{\theta_i\,|\,i\in{\cal I}_{\rm I}\}$: the set of eigenvalues of $J^{\rm I}$,

$\{\theta_i\,|\,i\in{\cal I}_{\rm II}\}$: the set of eigenvalues of $J^{\rm II}$,

$\{\theta_i\,|\,i\in{\cal I}_{\rm III}\}$: the set of eigenvalues of $J^{\rm III}$.

We will evaluate the eigenvalues of $J^{\rm I}$, $J^{\rm II}$ and $J^{\rm III}$ as follows;
\subsubsection{Eigenvalues of $J^{\rm I}$}
\label{subsubsec:J1}
Let $J^{\rm I}_{i'i}$ be the $(i',i)$ component of $J^{\rm I}$, then by (\ref{eqn:theorem1-1}),
\begin{align}
J^{\rm I}_{i'i}=\delta_{i'i}+\lambda_i^\ast D_{i',i}^\ast,\ i',i\in{\cal I}_{\rm I}.\label{eqn:J1seibun}
\end{align}
Let $I\in{\mathbb R}^{m_1\times m_1}$ denote the identity matrix and define $B\equiv I-J^{\rm I}$. Let $B_{i'i}$ be the $(i',i)$ component of $B$, then from (\ref{eqn:J1seibun}),
\begin{align}
B_{i'i}&=-\lambda^\ast_iD^\ast_{i',i}\\[1mm]
&=\lambda^\ast_i\ds\sum_{j=1}^n\ds\frac{P_j^{i'}P_j^i}{Q_j^\ast},\,i',i\in{\cal I}_{\rm I}.\label{eqn:componentofB}
\end{align}
Let $\{\beta_i\,|\,i\in{\cal I}_{\rm I}\}$ be the set of eigenvalues of $B$, then we have $\theta_i=1-\beta_i,\,i\in{\cal I}_{\rm I}$. In order to calculate the eigenvalues of $B$, we will define the following matrices. Similar calculations are performed in \cite{yu}.

Let us define
\begin{align}
\Phi_1&\equiv\begin{pmatrix}P^1\\\vdots\\P^{m_1}\end{pmatrix}\in\mathbb {R}^{m_1\times n},\\[3mm]
\Gamma&\equiv\left(-D_{i',i}^\ast\right)=\left(\ds\sum_{j=1}^n\ds\frac{P_j^{i'}P_j^i}{Q_j^\ast}\right)\in\mathbb {R}^{m_1\times m_1},\\[3mm]
\Lambda&\equiv{\rm diag}\left(\lambda_1^\ast,\cdots,\lambda_{m_1}^\ast\right)\in\mathbb {R}^{m_1\times m_1},\label{eqn:diagonalLambda}
\end{align}
where (\ref{eqn:diagonalLambda}) is the diagonal matrix with diagonal components $\lambda_1^\ast,\cdots,\lambda_{m_1}^\ast$. Furthermore,
\begin{align}
\sqrt{\Lambda}&\equiv{\rm diag}\left(\sqrt{\lambda_1^\ast},\cdots,\sqrt{\lambda_{m_1}^\ast}\right)\in\mathbb {R}^{m_1\times m_1},\\
\Omega&\equiv{\rm diag}\left((Q_1^\ast)^{-1},\cdots,(Q_n^\ast)^{-1}\right)\in\mathbb {R}^{n\times n},\\
\sqrt\Omega&\equiv{\rm diag}\left((Q_1^\ast)^{-1/2},\cdots,(Q_n^\ast)^{-1/2}\right)\in\mathbb {R}^{n\times n}.
\end{align}
Then, we have, by calculation,
\begin{align}
\sqrt\Lambda B\sqrt\Lambda^{-1}&=\sqrt\Lambda\Gamma\sqrt\Lambda\\
&=\sqrt\Lambda\Phi_1\Omega\ {^t}\Phi_1\ {^t}\sqrt\Lambda\\
&=\sqrt\Lambda\Phi_1\sqrt\Omega\ {^t}\sqrt\Omega\ {^t}\Phi_1\ {^t}\sqrt\Lambda\\
&=\sqrt\Lambda\Phi_1\sqrt\Omega\ {^t}\!\left(\sqrt\Lambda\Phi_1\sqrt\Omega\right).\label{eqn:L-1BL}
\end{align}

From (\ref{eqn:m1definition}), $\sqrt\Lambda$ is a regular matrix and from the assumption (\ref{eqn:rankmdefinition}), ${\rm rank}\,\Phi_1=m_1$. Therefore, by $m_1\leq m\leq n$, we have ${\rm rank}\,\sqrt\Lambda\Phi_1\sqrt\Omega=m_1$, and thus from (\ref{eqn:L-1BL}), $\sqrt\Lambda B\sqrt\Lambda^{-1}$ is symmetric and positive\ definite. In particular, all the eigenvalues $\beta_1,\cdots,\beta_{m_1}$ of $B$ are positive. Without loss of generality, let $\beta_1\geq\cdots\geq\beta_{m_1}>0$.
By (\ref{eqn:componentofB}), every component of $B$ is non-negative and by Lemma \ref{lem:rowsumofJis0}, every row sum of $B$ is equal to 1, hence by the Perron-Frobenius theorem
\begin{align}
1=\beta_1\geq\beta_2\geq\cdots\geq\beta_{m_1}>0.
\end{align}
Because $\theta_i=1-\beta_i,\,i\in{\cal I}_{\rm I}$, we have
\begin{align}
0=\theta_1\leq\theta_2\leq\cdots\leq\theta_{m_1}<1,
\end{align}
therefore,
\begin{theorem}
\label{the:2}
The eigenvalues of $J^{\rm I}$ satisfy
\begin{align}
0\leq\theta_i<1,\,i\in{\cal I}_{\rm I}.\label{eqn:J1nokoyuchi}
\end{align}
\end{theorem}

\subsubsection{Eigenvalues of $J^{\rm II}$}
\label{subsubsec:J2}
From (\ref{eqn:J1AJ2}),\,(\ref{eqn:Jstructure2}), we have
\begin{theorem}
\label{the:3}
The eigenvalues of $J^{\rm II}$ satisfy
\begin{align}
\theta_i=1,\,i\in{\cal I}_{\rm II}.\label{eqn:J2nokoyuchi}
\end{align}
\end{theorem}

\subsubsection{Eigenvalues of $J^{\rm III}$}
\label{subsubsec:J3}
From (\ref{eqn:J1AJ2}),\,(\ref{eqn:Jstructure3}), we have
\begin{theorem}
\label{the:4}
The eigenvalues of $J^{\rm III}$ are $\theta_i=e^{D^\ast_i-C},\,D^\ast_i<C,\,i\in{\cal I}_{\rm III}$, hence
\begin{align}
0<\theta_i<1,\,i\in{\cal I}_{\rm III}.\label{eqn:J3nokoyuchi}
\end{align}
\end{theorem}

\begin{remark}
\rm From the above consideration, we know that all the eigenvalues of the Jacobian matrix $J(\bm\lambda^\ast)$ are real.
\end{remark}

\section{On convergence speed}
We obtained in Theorems \ref{the:2},\,\ref{the:3},\,\ref{the:4}, the evaluation for the eigenvalues of $J(\bm\lambda^\ast)$. Let $\theta_{\rm max}\equiv\max_{i\in{\cal I}}\theta_i$ be the maximum eigenvalue of $J(\bm\lambda^\ast)$, then by Theorems \ref{the:2},\,\ref{the:3},\,\ref{the:4}, we have $0\leq\theta_{\rm max}<1$ if ${\cal I}_{\rm II}$ is empty and $\theta_{\rm max}=1$ if ${\cal I}_{\rm II}$ is not empty. In the following, we will see that $\bm\lambda^N\to\bm\lambda^\ast$ or $\bm\mu^N\to\bm0$ is the exponential convergence if $0\leq\theta_{\rm max}<1$, and the $1/N$ order convergence if $\theta_{\rm max}=1$.
\subsection{Convergence speed in case of $0\leq\theta_{\rm max}<1$}
\begin{theorem}
\label{the:5}
Suppose that the maximum eigenvalue $\theta_{\rm max}$ of the Jacobian matrix $J(\bm\lambda^\ast)$ satisfies $0\leq\theta_{\rm max}<1$. Then, for any $\theta$ with $\theta_{\rm max}<\theta<1$, there exist $\delta>0$ and $K>0$, such that for arbitrary initial vector $\bm\lambda^0$ with $\|\bm\lambda^0-\bm\lambda^\ast\|<\delta$, we have
\begin{align}
\|\bm\mu^N\|=\|\bm\lambda^N-\bm\lambda^\ast\|<K\theta^N,\,N=0,1,\cdots,
\end{align}
i.e., the convergence is exponential, where $\theta^N$ denotes the $N$th power of $\theta$.
\end{theorem}
{\bf Proof:}
See Appendix \ref{sec:proooftheorem4}.\hfill$\blacksquare$

%\medskip

%\begin{remark}
%In \cite{ari},\cite{yu}, they consider the maximum eigenvalue (or spectral radius) $\theta_{\rm max}$ as the convergence rate, however, the convergence rate must be slightly larger than $\theta_{\rm max}$, as shown in this theorem.
%\end{remark}

%
\subsection{Convergence speed in case of $\theta_{\rm max}=1$}
In the case of $\theta_{\rm max}=1$, Theorem \ref{the:5} cannot be applied, i.e., the convergence $\bm\mu^N\to\bm0$ is not determined only by the Jacobian matrix, but it is necessary to investigate the Hessian matrix of the second order term of the Taylor expansion.
\subsection{On Hessian matrix}
In the previous studies, say, \cite{ari},\cite{mat},\cite{yu}, the Jacobian matrix is considered but the Hessian matrix is not. Let us calculate the components (\ref{eqn:Hesseseibun}) of the Hessian matrix of the function $F_i,\,i=1,\cdots,m$, at $\bm\lambda=\bm\lambda^\ast$. Define $D^\ast_{i,i',i''}\equiv\partial^2D_i/\partial\lambda_{i'}\partial\lambda_{i{''}}|_{\bm\lambda=\bm\lambda^\ast}.$ We have
\begin{theorem}
\label{the:6}
\begin{align}
\label{eqn:theorem5-1}
&\left.\ds\frac{\partial^2F_i}{\partial\lambda_{i'}\partial\lambda_{i''}}\right|_{\bm\lambda=\bm\lambda^\ast}=e^{D_i^\ast-C}\Big[(1-e^{D_{i'}^\ast-C}+D_{i,i'}^\ast)(\delta_{ii''}+\lambda_i^\ast(1-e^{D_{i''}^\ast-C}))\nonumber\\[3mm]
&\ \ +(1-e^{D_{i''}^\ast-C}+D_{i,i''}^\ast)(\delta_{ii'}+\lambda_i^\ast(1-e^{D_{i'}^\ast-C}))\nonumber\\
&\ \ +\lambda_i^\ast\Big(D_{i,i'}^\ast D_{i,i''}^\ast+D_{i,i',i''}^\ast+D_{i',i''}^\ast-e^{D_{i'}^\ast-C}D_{i',i''}^\ast-e^{D_{i''}^\ast-C}D_{i',i''}^\ast-\ds\sum_{k=1}^{m_1}\lambda_k^\ast D_{k,i'}^\ast D_{k,i''}^\ast\Big)\Big],\nonumber\\
&\ \ i,i',i''\in{\cal I}.
\end{align}
Especially, if ${\cal I}_{\rm III}$ is empty, then for $i\in{\cal I}_{\rm II}$,
\begin{align}
&\left.\ds\frac{\partial^2F_i}{\partial\lambda_{i'}\partial\lambda_{i''}}\right|_{\bm\lambda=\bm\lambda^\ast}=\delta_{ii'}D_{i,i''}^\ast+\delta_{ii''}D_{i,i'}^\ast,\,i',i''\in{\cal I},\label{eqn:theorem5-3}
\end{align}
which is a relatively simple form.
\end{theorem}
{\bf Proof:}
See Appendix \ref{sec:proofoftheorem5}.\hfill$\blacksquare$

\section{Convergence speed in case of $m=3$ and $n$ is arbitrary}
\label{sec:m3narbitray}
In Theorem \ref{the:6}, the Hessian matrix is very complicated, thus it is difficult to investigate arbitrary channel matrix. Therefore, in this section, we will consider a special case, i.e., $m=3$ and $n$ is arbitrary. For $m=3$, without loss of generality, we have the following exhaustive classification.
\begin{itemize}
\item[(i)] $\lambda^\ast_1>0,\lambda^\ast_2>0,\lambda^\ast_3>0,$
\item[(ii)] $\lambda^\ast_1>0,\lambda^\ast_2>0,\lambda^\ast_3=0,D^\ast_3=C,$
\item[(iii)] $\lambda^\ast_1>0,\lambda^\ast_2>0,\lambda^\ast_3=0,D^\ast_3<C.$
\end{itemize}
(i) is the case of ``acute triangle'' in Example \ref{exa:1}. We have ${\cal I}_{\rm I}={\cal I}$, ${\cal I}_{\rm II}={\cal I}_{\rm III}=\emptyset$, thus by (\ref{eqn:J1AJ2}),\,(\ref{eqn:Jstructure1}),
\begin{align}
J(\bm\lambda^\ast)=J^{\rm I}.\label{eqn:3by3Jacobimatrix1}
\end{align}
By Theorem \ref{the:2}, we have $0\leq\theta_{\rm max}<1$ then, by Theorem \ref{the:5} the convergence $\bm\mu^N\to\bm0$ is exponential. 

Skipping (ii), let us consider (iii) first. (iii) is the case of ``obtuse triangle'' in Example \ref{exa:3}. We have ${\cal I}_{\rm I}=\{1,2\}$, ${\cal I}_{\rm II}=\emptyset$, ${\cal I}_{\rm III}=\{3\}$, thus by (\ref{eqn:J1AJ2}),\,(\ref{eqn:Jstructure3}),
\begin{align}
&J(\bm\lambda^\ast)=\begin{pmatrix}
\,J^{\rm I} & O\,\\
\,\ast & J^{\rm III}\,
\end{pmatrix},\label{eqn:3by3Jacobimatrix3}\\
&\ J^{\rm I}\in{\mathbb R}^{2\times2},\\
&\ J^{\rm III}=e^{D^\ast_3-C},\,0<J^{\rm III}<1.
\end{align}
By Theorems \ref{the:2},\,\ref{the:4}, we have $0<\theta_{\rm max}<1$, then by Theorem \ref{the:5}, the convergence $\bm\mu^N\to\bm0$ is exponential.

The rest is (ii), which is the case of ``right triangle'' in Example \ref{exa:2}. In this case, we have ${\cal I}_{\rm I}=\{1,2\}$, ${\cal I}_{\rm II}=\{3\}$, ${\cal I}_{\rm III}=\emptyset$, thus by (\ref{eqn:J1AJ2}),\,(\ref{eqn:Jstructure2}),
\begin{align}
&J(\bm\lambda^\ast)=\begin{pmatrix}
\,J^{\rm I} & O\,\\
\,\ast & J^{\rm II}\,
\end{pmatrix},\label{eqn:3by3Jacobimatrix2}\\
&\ J^{\rm I}\in{\mathbb R}^{2\times2},\\
&\ J^{\rm II}=1.
\end{align}
By Theorems \ref{the:2},\,\ref{the:3}, $\theta_{\rm max}=1$, thus we cannot apply Theorem \ref{the:5}. For the analysis of the convergence speed, we will investigate the Hessian matrix in the second order term of the Taylor expansion.
\subsection{Convergence of $1/N$ order }
We will investigate the convergence speed of $\bm\mu^N\to\bm0$ in the case (ii) above and prove that it is the convergence of the $1/N$ order.

By (\ref{eqn:theorem1-1}) in Theorem \ref{the:1} and (\ref{eqn:theorem5-3}) in Theorem \ref{the:6}, we have $J(\bm\lambda^\ast)$ and $H_3(\bm\lambda^\ast)$ as
\begin{align}
J(\bm\lambda^\ast)&=\begin{pmatrix}
\,1+\lambda^\ast_1D^\ast_{1,1} & \lambda^\ast_2D^\ast_{1,2} & 0\,\\
\,\lambda^\ast_1D^\ast_{1,2} & 1+\lambda^\ast_2D^\ast_{2,2} & 0\,\\
\,\lambda^\ast_1D^\ast_{1,3} & \lambda^\ast_2D^\ast_{2,3} & 1\,
\end{pmatrix},\\
H_3(\bm\lambda^\ast)&=\begin{pmatrix}
\,0 & 0 & D^\ast_{1,3}\,\\
\,0 & 0 & D^\ast_{2,3}\,\\
\,D^\ast_{1,3} & D^\ast_{2,3} & 2D^\ast_{3,3}\,
\end{pmatrix}.
\end{align}
$H_1(\bm\lambda^\ast)$ and $H_2(\bm\lambda^\ast)$ do not affect directly on the convergence speed.

Now, we show some properties of 
\begin{align}
D^\ast_{i',i}=-\ds\sum_{j=1}^n\ds\frac{P^{i'}_jP^i_j}{Q^\ast_j},\,i',i=1,2,3,
\end{align}
defined by (\ref{eqn:Diidefinition}),\,(\ref{eqn:lem3-2}). We have
\begin{align}
&D^\ast_{i',i}=D^\ast_{i,i'},\,i',i=1,2,3,\label{eqn:property1}\\
&D^\ast_{i',i}\leq0,\,i',i=1,2,3,\label{eqn:property2}\\
%\end{align}
%
%\begin{align}
&\lambda^\ast_1D^\ast_{1,i}+\lambda^\ast_2D^\ast_{2,i}=-\ds\sum_{j=1}^nP^i_j\ds\sum_{i'=1}^2\ds\frac{\lambda^\ast_{i'}P^{i'}_j}{Q^\ast_j}\nonumber\\
&\hspace{24mm}=-1,\,i=1,2,3.\label{eqn:property3}
\end{align}

Let us consider the first order term
\begin{align}
\bm\mu^{N+1}=\bm\mu^NJ(\bm\lambda^\ast)\label{eqn:taylor1ji}
\end{align}
of the Taylor expansion (\ref{eqn:Taylortenkai3}). See also (\ref{eqn:muislambdaminuslambda}),\,(\ref{eqn:musumiszero}). The representation by components of (\ref{eqn:taylor1ji}) is 
\begin{align}
(\mu^{N+1}_1,\mu^{N+1}_2,\mu^{N+1}_3)=(\mu^N_1,\mu^N_2,\mu^N_3)\begin{pmatrix}
\,1+\lambda^\ast_1D^\ast_{1,1} & \lambda^\ast_2D^\ast_{1,2} & 0\,\\
\lambda^\ast_1D^\ast_{1,2} & 1+\lambda^\ast_2D^\ast_{2,2} & 0\\
\,\lambda^\ast_1D^\ast_{1,3} & \lambda^\ast_2D^\ast_{2,3} & 1\,
\end{pmatrix}.
\end{align}
Then, by calculation
\begin{align}
\mu_1^{N+1}&=(1+\lambda^\ast_1D^\ast_{1,1})\mu_1^N+\lambda^\ast_1D^\ast_{1,2}\,\mu^N_2+\lambda^\ast_1D^\ast_{1,3}\,\mu^N_3,\label{eqn:mu1}\\
\mu_2^{N+1}&=\lambda^\ast_2D^\ast_{1,2}\,\mu^N_1+(1+\lambda^\ast_2D^\ast_{2,2})\mu^N_2+\lambda^\ast_2D^\ast_{2,3}\,\mu^N_3,\label{eqn:mu2}\\
\mu_3^{N+1}&=\mu_3^N.
\end{align}
Substituting $\mu^N_3=-\mu^N_1-\mu^N_2$ into (\ref{eqn:mu1}),\,(\ref{eqn:mu2}),
\begin{align}
\mu_1^{N+1}&=(1+\lambda^\ast_1D^\ast_{1,1}-\lambda^\ast_1D^\ast_{1,3})\mu_1^N+(\lambda^\ast_1D^\ast_{1,2}-\lambda^\ast_1D^\ast_{1,3})\mu^N_2,\label{eqn:muhat1}\\
\mu_2^{N+1}&=(\lambda^\ast_2D^\ast_{1,2}-\lambda^\ast_2D^\ast_{2,3})\mu^N_1+(1+\lambda^\ast_2D^\ast_{2,2}-\lambda^\ast_2D^\ast_{2,3})\mu^N_2.\label{eqn:muhat2}
\end{align}
By defining
\begin{align}
&\hat{\bm\mu}^N\equiv(\mu^N_1,\,\mu^N_2),\\
&\hat{J}(\bm\lambda^\ast)\equiv\begin{pmatrix}
\,1+\lambda^\ast_1D^\ast_{1,1}-\lambda^\ast_1D^\ast_{1,3} & \lambda^\ast_2D^\ast_{1,2}-\lambda^\ast_2D^\ast_{2,3}\\
\lambda^\ast_1D^\ast_{1,2}-\lambda^\ast_1D^\ast_{1,3} & 1+\lambda^\ast_2D^\ast_{2,2}-\lambda^\ast_2D^\ast_{2,3}\,
\end{pmatrix},\label{eqn:hatJ}
\end{align}
(\ref{eqn:muhat1}) and (\ref{eqn:muhat2}) become
\begin{align}
\hat{\bm\mu}^{N+1}=\hat{\bm\mu}^N\hat{J}(\bm\lambda^\ast).\label{eqn:jacobi}
\end{align}
Let us calculate the eigenvalues and right eigenvectors of $\hat{J}(\bm\lambda^\ast)$. In the following calculation, (\ref{eqn:property3}) is often used. The characteristic polynomial $\varphi_{\hat{J}(\bm\lambda^\ast)}(\eta)\equiv\det\left(\hat{J}(\bm\lambda^\ast)-\eta I\right)$ of $\hat{J}(\bm\lambda^\ast)$ is%
\begin{align*}
&\varphi_{\hat{J}(\bm\lambda^\ast)}(\eta)\\
&=\det\begin{pmatrix}\,1+\lambda^\ast_1D^\ast_{1,1}-\lambda^\ast_1D^\ast_{1,3}-\eta & \hspace{-3mm}\lambda^\ast_2D^\ast_{1,2}-\lambda^\ast_2D^\ast_{2,3}\\
\lambda^\ast_1D^\ast_{1,2}-\lambda^\ast_1D^\ast_{1,3} & \hspace{-3mm}1+\lambda^\ast_2D^\ast_{2,2}-\lambda^\ast_2D^\ast_{2,3}-\eta\,\end{pmatrix}\\
& {\rm (Add\ the\ 2nd\ column\ to\ the\ 1st\ column\ to\ have)}\\
&=\det\begin{pmatrix}\,1-\eta & \lambda^\ast_2D^\ast_{1,2}-\lambda^\ast_2D^\ast_{2,3}\\
1-\eta & 1+\lambda^\ast_2D^\ast_{2,2}-\lambda^\ast_2D^\ast_{2,3}-\eta\,\end{pmatrix}\\
& {\rm (Add\ (-1)\times the\ 1st\ row\ to\ the\ 2nd\ row\ to\ have)}\\
&=\det\begin{pmatrix}\,1-\eta & \lambda^\ast_2D^\ast_{1,2}-\lambda^\ast_2D^\ast_{2,3}\\
0 & 1+\lambda^\ast_2D^\ast_{2,2}-\lambda^\ast_2D^\ast_{1,2}-\eta\,\end{pmatrix}\\
&=\det\begin{pmatrix}\,1-\eta & \lambda^\ast_2D^\ast_{1,2}-\lambda^\ast_2D^\ast_{2,3}\\
0 & -D^\ast_{1,2}-\eta\,\end{pmatrix}\\
&=(\eta+D^\ast_{1,2})(\eta-1).
\end{align*}
Thus, the eigenvalues of $\hat{J}(\bm\lambda^\ast)$ are $\eta_1\equiv-D^\ast_{1,2}$ and $\eta_2\equiv1$.
\begin{lemma}
\label{lem:4}
$0\leq\eta_1<1$.
\end{lemma}
{\bf Proof:} First, $\eta_1\geq0$ by (\ref{eqn:property2}). Next, if $-D^\ast_{1,2}<-D^\ast_{2,2}$, then by (\ref{eqn:property3}), $1=\lambda^\ast_1(-D^\ast_{1,2})+\lambda^\ast_2(-D^\ast_{2,2})>\lambda^\ast_1(-D^\ast_{1,2})+\lambda^\ast_2(-D^\ast_{1,2})=-D^\ast_{1,2}$, which proves $\eta_1=-D^\ast_{1,2}<1$. Thus, we will prove $-D^\ast_{1,2}<-D^\ast_{2,2}$. This inequality is equivalent to
\begin{align}
\ds\sum_{j=1}^n\frac{(P^2_j)^2}{Q^\ast_j}>\ds\sum_{j=1}^n\frac{P^1_jP^2_j}{Q^\ast_j}\label{eqn:lem1-1}
\end{align}
by (\ref{eqn:Diidefinition}),\,(\ref{eqn:lem3-2}). We will prove (\ref{eqn:lem1-1}).

\begin{figure}[t]
\begin{center}
\begin{overpic}[width=8cm]{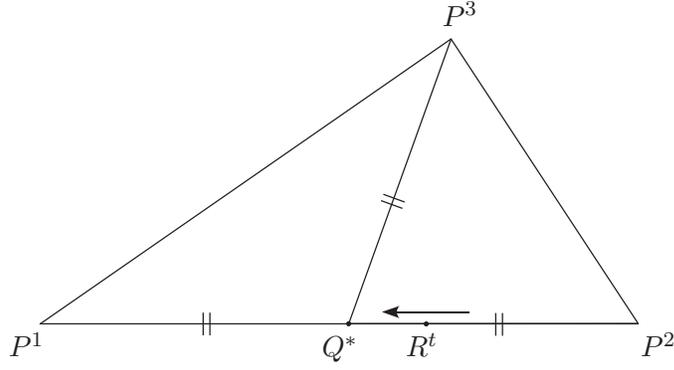}
%\begin{overpic}[width=7cm,grid]{./figure/fig3.eps}
\put(-5,-3){$P^1$}
\put(100,-3){$P^2$}
\put(67,52){$P^3$}
\put(47,-3){$Q^\ast$}
\put(61,-3){$R^t$}
\end{overpic}
\vspace{1mm}
\caption{Figure for the proof of Lemma \ref{lem:4}}
\label{fig:6}
\end{center}
\end{figure}

Let $R^t$ be a point on the line segment $P^1P^2$ moving from $P^2$ to $P^1$, i.e.,
\begin{align}
R^t\equiv(1-t)P^2+tP^1,\,0\leq t\leq1,\label{eqn:lem1-2}
\end{align}
see Fig.\ref{fig:6}. Write $R^t$ by components as $R^t=(R^t_1,\cdots,R^t_n)$.
%
%\vspace{-8mm}
Define a function $g(t)$ by
\begin{align}
g(t)\equiv D(P^2\|R^t)=\ds\sum_{j=1}^nP^2_j\log\ds\frac{P^2_j}{R^t_j}.\label{eqn:lem1-3}
\end{align}
Then, 
\begin{align}
g'(t)=\ds\sum_{j=1}^n\ds\frac{(P^2_j)^2}{R^t_j}-\ds\sum_{j=1}^n\ds\frac{P^1_jP^2_j}{R^t_j},\label{eqn:lem1-5}
\end{align}
and
\begin{align}
g''(t)=\ds\sum_{j=1}^nP^2_j\ds\frac{\left(P^2_j-P^1_j\right)^2}{\left(R^t_j\right)^2}>0.\label{eqn:gtwodash}
\end{align}
From (\ref{eqn:lem1-5}) and $R^0=P^2$, $g'(0)=0$. From (\ref{eqn:gtwodash}), $g'(t)$ is monotonically increasing, thus $g'(t)>g'(0)=0,\,0<t\leq1$. Since $R^{\lambda^\ast_1}=\lambda^\ast_2P^2+\lambda^\ast_1P^1=Q^\ast$, substituting $t=\lambda^\ast_1$ into (\ref{eqn:lem1-5}), we obtain
\begin{align}
0<g'(\lambda^\ast_1)=\ds\sum_{j=1}^n\ds\frac{(P^2_j)^2}{Q^\ast_j}-\ds\sum_{j=1}^n\ds\frac{P^1_jP^2_j}{Q^\ast_j},\label{eqn:lem1-6}
\end{align}
which proves (\ref{eqn:lem1-1}).\hfill$\blacksquare$

\medskip

Next, we will calculate a right eigenvector $\bm a=\begin{pmatrix}a_1\\ a_2\end{pmatrix}$ of $\hat{J}(\bm\lambda^\ast)$ for the eigenvalue $\eta_1=-D^\ast_{1,2}$. The equation
\begin{align}
\hat{J}(\bm\lambda^\ast)\bm a=\eta_1\bm a\label{eqn:eigenvectorcalculation01}
\end{align}
is written by components as
\begin{align}
\begin{pmatrix}\,1+\lambda^\ast_1D^\ast_{1,1}-\lambda^\ast_1D^\ast_{1,3} & \lambda^\ast_2D^\ast_{1,2}-\lambda^\ast_2D^\ast_{2,3}\\
\lambda^\ast_1D^\ast_{1,2}-\lambda^\ast_1D^\ast_{1,3} & 1+\lambda^\ast_2D^\ast_{2,2}-\lambda^\ast_2D^\ast_{2,3}\,\end{pmatrix}\begin{pmatrix}a_1\\ a_2\end{pmatrix}=-D^\ast_{1,2}\begin{pmatrix}a_1\\ a_2\end{pmatrix}.\label{eqn:eigenvectorcalculation02}
\end{align}
From (\ref{eqn:eigenvectorcalculation02}),\,(\ref{eqn:property1}),\,(\ref{eqn:property3}), we have, by calculation
\begin{align}
\lambda^\ast_1(D^\ast_{1,2}-D^\ast_{1,3})a_1+\lambda^\ast_2(D^\ast_{1,2}-D^\ast_{2,3})a_2=0.\label{eqn:eigenvectorcalculation03}
\end{align}
By defining
\begin{align}
\tau_1\equiv\lambda^\ast_1(D^\ast_{1,2}-D^\ast_{1,3}),\ \tau_2\equiv\lambda^\ast_2(D^\ast_{1,2}-D^\ast_{2,3}),\label{eqn:eigenvectorcalculation04}
\end{align}
(\ref{eqn:eigenvectorcalculation03}) is written as
\begin{align}
\tau_1a_1+\tau_2a_2=0.\label{eqn:eigenvectorcalculation05}
\end{align}

Now, by (\ref{eqn:property3}) and Lemma \ref{lem:4}, we have
\begin{align}
\tau_1+\tau_2=1+D^\ast_{1,2}>0.\label{eqn:eigenvectorcalculation06}
\end{align}
We notice that $a_1\neq a_2$. In fact, if $a_1=a_2$, we have $a_1=a_2\neq0$ because $\bm a$ is an eigenvector, then (\ref{eqn:eigenvectorcalculation05}) and (\ref{eqn:eigenvectorcalculation06}) contradict each other. Hence, we can impose
\begin{align}
a_1-a_2=1\label{eqn:eigenvectorcalculation07}
\end{align}
as a normalizing condition of the eigenvector. By solving (\ref{eqn:eigenvectorcalculation05}) and (\ref{eqn:eigenvectorcalculation07}), we have
\begin{align}
&a_1=\dfrac{\tau_2}{\tau_1+\tau_2}=\dfrac{\lambda^\ast_2(D^\ast_{1,2}-D^\ast_{2,3})}{1+D^\ast_{1,2}},\\
&a_2=-\dfrac{\tau_1}{\tau_1+\tau_2}=-\dfrac{\lambda^\ast_1(D^\ast_{1,2}-D^\ast_{1,3})}{1+D^\ast_{1,2}}.\label{eqn:eigenvectorcalculation08}
\end{align}

Multiplying the both sides of (\ref{eqn:jacobi}) by $\bm a$ from the right, we have
\begin{align}
\hat{\bm\mu}^{N+1}\bm a&=\hat{\bm\mu}^N\hat{J}(\bm\lambda^\ast)\bm a\nonumber\\
&=\eta_1\hat{\bm\mu}^N\bm a\nonumber\\
&=\cdots\nonumber\\
&=\left(\eta_1\right)^{N+1}\hat{\bm\mu}^0\bm a.
\end{align}
Putting $K\equiv\hat{\bm\mu}^0\bm a$, we have
\begin{align}
\hat{\bm\mu}^N\bm a=K\left(\eta_1\right)^N,
\end{align}
and by components
\begin{align}
a_1\mu^N_1+a_2\mu^N_2=K\left(\eta_1\right)^N.\label{eqn:eigenvectorcalculation09}
\end{align}
Then, from (\ref{eqn:eigenvectorcalculation09}) and $\mu^N_1+\mu^N_2=-\mu^N_3$, we have
\begin{align}
\mu_1^N&=a_2\mu^N_3+K\left(\eta_1\right)^N,\label{eqn:eigenvectorcalculation10}\\[1mm]
\mu_2^N&=-a_1\mu^N_3-K\left(\eta_1\right)^N.\label{eqn:eigenvectorcalculation11}
\end{align}
Defining $b_1\equiv-a_2,\,b_2\equiv a_1$, we obtain the following results;
\begin{align}
\mu_1^N&=-b_1\mu^N_3+K\left(\eta_1\right)^N,\label{eqn:mun1}\\[1mm]
\mu_2^N&=-b_2\mu^N_3-K\left(\eta_1\right)^N,\label{eqn:mun2}
\end{align}
where
\begin{align}
b_1&\equiv\frac{\lambda^\ast_1(D^\ast_{1,2}-D^\ast_{1,3})}{1+D^\ast_{1,2}},\label{eqn:b1}\\[1mm]
b_2&\equiv\frac{\lambda^\ast_2(D^\ast_{1,2}-D^\ast_{2,3})}{1+D^\ast_{1,2}}.\label{eqn:b2}
\end{align}
We have $b_1+b_2=1$.

\begin{remark}
\rm As for the eigenvalue $\eta_2=1$, an eigenvector is $\begin{pmatrix}\,1\,\\1\end{pmatrix}$ and $\hat{J}(\bm\lambda^\ast)\begin{pmatrix}\,1\,\\1\end{pmatrix}=1\begin{pmatrix}\,1\,\\1\end{pmatrix}$ only shows a trivial relation because of (\ref{eqn:property3}).
\end{remark}

\begin{remark}
\label{rem:5}
\rm We obtained (\ref{eqn:mun1})-(\ref{eqn:b2}) by regarding (\ref{eqn:taylor1ji}) holds exactly. Actually, (\ref{eqn:taylor1ji}) holds approximately if $N$ is sufficiently large and the second and higher order terms of Taylor expansion are sufficiently small. Therefore, (\ref{eqn:mun1})-(\ref{eqn:b2}) also hold approximately. In particular, the approximate value for $\eta_1$ in Lemma \ref{lem:4} is considered to be smaller than $1$. Refer the proof of Theorem \ref{the:5}.
\end{remark}

\bigskip

Now, consider the third component of the Taylor expansion (\ref{eqn:Taylortenkai3});
\begin{align}
&\mu^{N+1}_3=\bm\mu^N\begin{pmatrix}\,0\,\\0\\1\end{pmatrix}+\ds\frac{1}{2!}\bm\mu^NH_3(\bm\lambda^\ast)\,{^t}\bm\mu^N+o\left(\|\bm\mu^N\|^2\right)\nonumber\\
&=\mu^N_3+\ds\frac{1}{2!}\left(\mu^N_1,\mu^N_2,\mu^N_3\right)
\begin{pmatrix}
0 & 0 & D^\ast_{1,3}\\
0 & 0 & D^\ast_{2,3}\\
\,D^\ast_{1,3} & D^\ast_{2,3} & 2D^\ast_{3,3}\,
\end{pmatrix}
\begin{pmatrix}\,\mu^N_1\,\\[1mm]\mu^N_2\\[1mm]\mu^N_3\end{pmatrix}\nonumber\\
&\ \ \ \ +o\left(\|\bm\mu^N\|^2\right)\nonumber\\
&=\mu^N_3+D^\ast_{1,3}\mu^N_1\mu^N_3+D^\ast_{2,3}\mu^N_2\mu^N_3+D^\ast_{3,3}\left(\mu^N_3\right)^2\nonumber\\
&\ \ \ \ +o\left(\|\bm\mu^N\|^2\right)\nonumber\\
&=\mu^N_3-(D^\ast_{1,3}b_1+D^\ast_{2,3}b_2-D^\ast_{3,3})\left(\mu^N_3\right)^2+o\left((\mu^N_3)^2\right),\label{eqn:thirdcomponent}
\end{align}
where the last equality is obtained by (\ref{eqn:mun1}),\,(\ref{eqn:mun2}). Defining
\begin{align}
&\rho\equiv D^\ast_{1,3}b_1+D^\ast_{2,3}b_2-D^\ast_{3,3}\label{eqn:rhodefinition}\\[1mm]
&=\lambda^\ast_1\ds\frac{D^\ast_{1,3}(D^\ast_{1,2}-D^\ast_{1,3})}{1+D^\ast_{1,2}}+\lambda^\ast_2\ds\frac{D^\ast_{2,3}(D^\ast_{1,2}-D^\ast_{2,3})}{1+D^\ast_{1,2}}-D^\ast_{3,3},
\end{align}
we have by (\ref{eqn:thirdcomponent}),\,(\ref{eqn:rhodefinition}),
\begin{align}
\mu^{N+1}_3=\mu^N_3-\rho\left(\mu^N_3\right)^2+o\left((\mu^N_3)^2\right).\label{eqn:mu3zenkashiki}
\end{align}

Now we assume
\begin{align}
\rho>0.\label{eqn:rhopositiveassumption}
\end{align}
If $\rho<0$, then the recurrence formula (\ref{eqn:mu3zenkashiki}) diverges, hence $\rho\geq0$ holds. Thus, the assumption (\ref{eqn:rhopositiveassumption}) is equivalent to $\rho\neq0$.
\begin{lemma}
\label{lem:5}
Consider the recurrence formula $(\ref{eqn:mu3zenkashiki})$. For a sufficiently small $\delta>0$ and any initial value $\mu^0_3$ with $0<\mu^0_3<\delta$, we have $\ds\lim_{N\to\infty}\mu^N_3=0$.
\end{lemma}
{\bf Proof:}
Consider the function $\mu-\rho\mu^2+o(\mu^2)$. If $\delta>0$ is sufficiently small, then for any $\mu$ with $0<\mu<\delta$, we have $\rho\mu+o(\mu)<1$ and $(\rho/2)\mu^2>o\left(\mu^2\right)$. Thus, for any initial value $\mu^0_3$ with $0<\mu^0_3<\delta$ we have $\mu^1_3=\mu^0_3\left(1-\rho\mu^0_3+o\left(\mu^0_3\right)\right)>0$ and $\mu^1_3<\mu^0_3-(\rho/2)\left(\mu^0_3\right)^2<\mu^0_3<\delta$. By mathematical induction, we have $0<\mu^N_3<\delta,\,N=0,1,\cdots$, hence
\begin{align}
0<\mu^{N+1}_3<\mu^N_3-\ds\frac{\rho}{2}(\mu^N_3)^2,\ N=0,1,\cdots.\label{eqn:zenkahutoushiki}
\end{align}
Since $0<\mu^{N+1}_3<\mu^N_3$ holds by (\ref{eqn:zenkahutoushiki}), there exists the limit $\mu^\infty_3\equiv\ds\lim_{N\to\infty}\mu^N_3\geq0$. Letting $N\to\infty$ in (\ref{eqn:zenkahutoushiki}), we have $\mu^\infty_3\leq\mu^\infty_3-(\rho/2)\left(\mu^\infty_3\right)^2$, which implies $\mu^\infty_3=0$.\hfill$\blacksquare$

%\bigskip

\begin{lemma}
\label{lem:6}
For a sufficiently small $\delta>0$ and any initial value $\mu^0_3$ with $0<\mu^0_3<\delta$, we have
\begin{align}
\ds\lim_{N\to\infty}N\mu^N_3=\ds\frac{1}{\rho}.\label{eqn:convergence1overrho}
\end{align}
\end{lemma}
{\bf Proof:} From (\ref{eqn:mu3zenkashiki}),
\begin{align}
\ds\frac{1}{\mu^{l+1}_3}-\ds\frac{1}{\mu^l_3}&=\ds\frac{1}{\mu^l_3-\rho\left(\mu^l_3\right)^2+o\left((\mu^l_3)^2\right)}-\ds\frac{1}{\mu^l_3}\\[2mm]
&=\ds\frac{\rho+o\left((\mu^l_3)^2\right)/(\mu^l_3)^2}{1-\rho\mu^l_3+o\left((\mu^l_3)^2\right)/|\mu^l_3|},\label{eqn:theorem3-1}
\end{align}
hence taking the arithmetic mean of the both sides of (\ref{eqn:theorem3-1}) for $l=0,1,\cdots,N-1$, 
\begin{align}
\ds\frac{1}{N}\ds\sum_{l=0}^{N-1}\left(\ds\frac{1}{\mu^{l+1}_3}-\ds\frac{1}{\mu^l_3}\right)=\ds\frac{1}{N}\ds\sum_{l=0}^{N-1}\ds\frac{\rho+o\left((\mu^l_3)^2\right)/(\mu^l_3)^2}{1-\rho\mu^l_3+o\left((\mu^l_3)^2\right)/|\mu^l_3|}.\label{eqn:therem3-2}
\end{align}
Applying the proposition that ``the arithmetic mean of a convergent sequence converges to the same limit as the original sequence'' \cite{ahl},\,p.37, to the right hand side of (\ref{eqn:therem3-2}), and further, by Lemma \ref{lem:5},
\begin{align*}
\lim_{N\to\infty}\ds\frac{1}{N}\left(\ds\frac{1}{\mu^N_3}-\ds\frac{1}{\mu^0_3}\right)&=\lim_{N\to\infty}\ds\frac{\rho+o\left((\mu^N_3)^2\right)/(\mu^N_3)^2}{1-\rho\mu^N_3+o\left((\mu^N_3)^2\right)/|\mu^N_3|}\\
&=\rho,
\end{align*}
which proves (\ref{eqn:convergence1overrho}).\hfill$\blacksquare$

\medskip

From (\ref{eqn:mun1}),\,(\ref{eqn:mun2}) and Lemma \ref{lem:6}, we have
\begin{theorem}
\label{the:7}
Let $m=3$ and $n$ be arbitrary. Suppose that the capacity achieving $\bm\lambda^\ast=(\lambda^\ast_1,\lambda^\ast_2,\lambda^\ast_3)$ satisfies $\lambda^\ast_1>0,\lambda^\ast_2>0,\lambda^\ast_3=0$ and $D^\ast_3=D(P^3\|\bm\lambda^\ast\Phi)=C$ $($see the case {\rm (ii)} at the first part of section $\ref{sec:m3narbitray})$, and further, $\rho>0$ in $(\ref{eqn:rhopositiveassumption})$. Then for $\bm\mu^N=\bm\lambda^N-\bm\lambda^\ast$ with $\bm\mu^N=(\mu^N_1,\mu^N_2,\mu^N_3)$, the convergence $\bm\mu^N\to\bm0$ is the $1/N$ order and we have
\begin{align}
&\ds\lim_{N\to\infty}N\mu^N_1=-\ds\frac{b_1}{\rho},\\[2mm]
&\ds\lim_{N\to\infty}N\mu^N_2=-\ds\frac{b_2}{\rho},\\[2mm]
&\ds\lim_{N\to\infty}N\mu^N_3=\ds\frac{1}{\rho},
\end{align}
where
$b_1=\ds\frac{\lambda^\ast_1(D^\ast_{1,2}-D^\ast_{1,3})}{1+D^\ast_{1,2}},\ b_2=\ds\frac{\lambda^\ast_2(D^\ast_{1,2}-D^\ast_{2,3})}{1+D^\ast_{1,2}},\\[2mm]
\rho=\lambda^\ast_1\ds\frac{D^\ast_{1,3}(D^\ast_{1,2}-D^\ast_{1,3})}{1+D^\ast_{1,2}}+\lambda^\ast_2\ds\frac{D^\ast_{2,3}(D^\ast_{1,2}-D^\ast_{2,3})}{1+D^\ast_{1,2}}-D^\ast_{3,3}$,\ and $D^\ast_{i',i}$ was defined by $(\ref{eqn:Diidefinition}),\,(\ref{eqn:lem3-2})$.
\end{theorem}

\subsection{Summary of Section \ref{sec:m3narbitray}}
We examined in this section the convergence speed of the Arimoto algorithm in the case that $m=3$ and $n$ is arbitrary. Based on the exhaustive classification (i), (ii), (iii) shown at the first part of section \ref{sec:m3narbitray}, in (i), (iii) the convergence is exponential, and in (ii) it is the $1/N$ order, under the assumption of $\rho>0$. In (ii), type II index in (\ref{eqn:Kuhn-Tucker2}) exists, therefore, under the assumption of $\rho>0$, we obtain the following equivalence;

\medskip

type II index exists $\Longleftrightarrow\theta_{\rm max}=1$ $\Longleftrightarrow$ the convergence is the $1/N$ order

\medskip

\noindent We conjecture that the same equivalence holds also in the case $m>3$.

\section{Numerical Evaluation}
Based on the analysis in the previous sections, we will evaluate numerically the convergence speed of the Arimoto algorithm for several channel matrices with $m=n=3$.

In Examples \ref{exa:4} and \ref{exa:5} below, we will investigate the exponential convergence in the case (i) in section \ref{sec:m3narbitray}, where the capacity achieving $\bm\lambda^\ast$ is in $\Delta({\cal X})^\circ$ (the interior of $\Delta({\cal X})$). In Example \ref{exa:5}, we will discuss how the convergence speed varies depending on the choice of initial input distribution $\bm\lambda^0$. Next, in Examples \ref{exa:6} and \ref{exa:7}, we will consider the $1/N$ order convergence in the case (ii). It will be confirmed that the convergence speed is accurately approximated by the limit values obtained in Theorem \ref{the:7}. In Example \ref{exa:8}, we will investigate the exponential convergence in the case (iii), where $\bm\lambda^\ast$ is on $\partial\Delta({\cal X})$ (the boundary of $\Delta({\cal X})$).

Here, in the exponential convergence, we will evaluate the values of the function
\begin{align}
L(N)\equiv-\ds\frac{1}{N}\log\|\bm\mu^N\|.
\end{align}
Based on the results of Theorem \ref{the:5}, i.e., $\|\bm\mu^N\|=\|\bm\lambda^N-\bm\lambda^\ast\|<K\theta^N$, $\theta\doteqdot\theta_{\rm max}$, we will compare $L(N)$ for large $N$ with $-\log\theta_{\rm max}$ or other values. 

On the other hand, in the $1/N$ order convergence, we will evaluate 
\begin{align}
N\bm\mu^N=(N\mu^N_1,N\mu^N_2,N\mu^N_3).
\end{align}
We will compare $N\bm\mu^N$ for large $N$ with the limit values obtained in Theorem \ref{the:7}.

\subsection{Case (i): exponential convergence where $\bm\lambda^\ast\in\Delta({\cal X})^\circ$}
\begin{example}
\label{exa:4}
\rm Consider the channel matrix $\Phi^{(1)}$ of (\ref{eqn:Phi1}), i.e.,
\begin{align}
\Phi^{(1)}=
\begin{pmatrix}
0.800 & 0.100 & 0.100\\
0.100 & 0.800 & 0.100\\
0.250 & 0.250 & 0.500
\end{pmatrix}.
\end{align}
We have
\begin{align}
\bm\lambda^\ast&=(0.431,0.431,0.138),\\
Q^\ast&=(0.422,0.422,0.156),\\
J(\bm\lambda^\ast)&=
\begin{pmatrix}
\,0.308 & -0.191 & -0.117\,\cr
\,-0.191 & 0.308 & -0.117\,\cr
\,-0.369 & -0.369 & 0.738\,\cr
\end{pmatrix}.
\end{align}
The eigenvalues of $J(\bm\lambda^\ast)$ are $(\theta_1,\theta_2,\theta_3)=(0.000,0.500,$ $0.855)$. Then, $\theta_{\rm max}=\theta_3=0.855$. If we choose $\bm\lambda^0=(1/3,1/3,1/3)$ as an initial distribution, then for $N=500$,
\begin{align}
L(500)=0.161\doteqdot-\log\theta_{\rm max}=0.157.\label{eqn:example4speedcomparison}
\end{align}
See Fig.\ref{fig:7}.
\begin{figure}[t]
\begin{center}
%\noindent
\begin{overpic}[width=8cm]{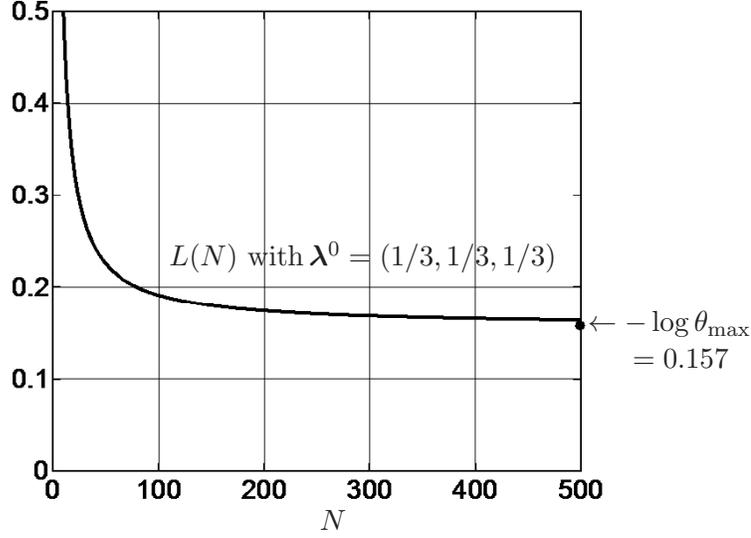}
%\begin{overpic}[width=7cm,grid]{./figure/absconv_c-0.00_2.eps}
\put(95.5,29){$\leftarrow-\log\theta_{\rm max}$}
\put(103,23){$=0.157$}
\put(51,-4){$N$}
\put(26,40){$L(N)\ \text{\rm with}\,\bm\lambda^0=(1/3,1/3,1/3)$}
\end{overpic}
\caption{Convergence of $L(N)$ in Example \ref{exa:4} with initial distribution $\bm\lambda^0=(1/3,1/3,1/3)$}
\label{fig:7}
\end{center}
\end{figure}
\end{example}

\begin{example}
\label{exa:5}
\rm Let us consider another channel matrix. Define
\begin{align}
\Phi^{(4)}\equiv
\begin{pmatrix}
\,0.793 & 0.196 & 0.011\,\\
0.196 & 0.793 & 0.011 \\
0.250 & 0.250 & 0.500
\end{pmatrix}.
\end{align}
We have
\begin{align}
\bm\lambda^\ast&=(0.352,0.352,0.296),\\
Q^\ast&=(0.422,0.422,0.156),\\
J(\bm\lambda^\ast)&=
\begin{pmatrix}
0.443 & -0.260 & -0.183\,\cr
-0.260 & 0.443 & -0.183\cr
\,-0.218 & -0.218 & 0.436
\end{pmatrix}.\label{eqn:example5J}
\end{align}
The eigenvalues of $J(\bm\lambda^\ast)$ are $(\theta_1,\theta_2,\theta_3)=(0.000,0.618,$ $0.702)$. Then, $\theta_{\rm max}=\theta_3=0.702$. Write the second largest eigenvalue as $\theta_{\rm sec}$, thus $\theta_{\rm sec}=\theta_2=0.618$.

We show in Fig.\ref{fig:8} the graph of $L(N)$ with initial distribution $\bar{\bm\lambda}^0\equiv(1/3,1/3,1/3)$ by solid line, and the graph with initial distribution $\bar{\bar{\bm\lambda}}^0\equiv(1/2,1/3,1/6)$ by dotted line.
\begin{figure}[t]
\begin{center}
%\noindent
\begin{overpic}[width=8cm]{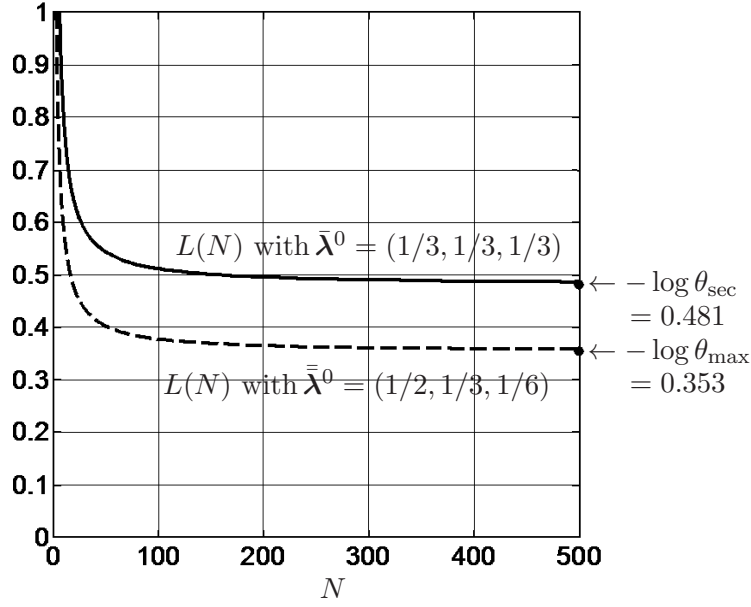}
%\begin{overpic}[width=7cm,grid]{./figure/example5.eps}
\put(27,53){$L(N)\ \text{\rm with}\,\bar{\bm\lambda}^0=(1/3,1/3,1/3)$}
\put(25,30){$L(N)\ \text{\rm with}\,\bar{\bar{\bm\lambda}}^0=(1/2,1/3,1/6)$}
\put(95.5,47){$\leftarrow-\log\theta_{\text{\rm sec}}$}
\put(102.5,41.5){$=0.481$}
\put(95.5,35.8){$\leftarrow-\log\theta_{\rm max}$}
\put(102.5,30.3){$=0.353$}
\put(51,-4){$N$}
\end{overpic}
\caption{Convergence of $L(N)$ in Example \ref{fig:5} with initial distribution $\bar{\bm\lambda}^0=(1/3,1/3,1/3)$ and $\bar{\bar{\bm\lambda}}^0=(1/2,1/3,1/6)$}
\label{fig:8}
\end{center}
\end{figure}
The larger $L(N)$ the faster the convergence, hence the convergence with $\bar{\bm\lambda}^0$ is faster than with $\bar{\bar{\bm\lambda}}^0$. The convergence speed varies depending on the choice of initial distribution. What kind of initial distribution yields faster convergence? We will investigate it below.

First, we consider the initial vector by $\bm\mu$ not by $\bm\lambda$, and define
\begin{align}
\bar{\bm\mu}^0&\equiv\bar{\bm\lambda}^0-\bm\lambda^\ast=(-0.019,-0.019,0.038),\label{eqn:initialmu0}\\
\bar{\bar{\bm\mu}}^0&\equiv\bar{\bar{\bm\lambda}}^0-\bm\lambda^\ast=(0.148,-0.019,-0.129).\label{eqn:initialbarmu0}
\end{align}

Similarly to Remark \ref{rem:5}, we will execute the following calculation by regarding $\bm\mu^{N+1}=\bm\mu^NJ(\bm\lambda^\ast),\,N=0,1,\cdots$ holds exactly.

Here, we will investigate for general $m,n$. We assume for simplicity that all the eigenvalues of $J(\bm\lambda^\ast)$ are different. Let $\bm\nu_{\rm max}$ be the left eigenvector of $J(\bm\lambda^\ast)$ for $\theta_{\rm max}$, and let $\bm\nu_{\rm max}^\perp$ be the orthogonal complement of $\bm\nu_{\rm max}$, i.e., $\bm\nu_{\rm max}^\perp\equiv\{\bm\mu\,|\,\bm\mu{^t}\bm\nu_{\rm max}=0\}.$

\begin{lemma}
\label{lem:thetasec}
If 
\begin{align}
\bm\mu^N\in\bm\nu_{\rm max}^\perp,\,N=0,1,\cdots,\label{eqn:innuperp}
\end{align}
then $\|\bm\mu^N\|<K\left(\theta_{\rm sec}\right)^N,\,K>0,\,N=0,1,\cdots$.
\end{lemma}
{\bf Proof:}
See Appendix \ref{sec:proofofthetasec}.\hfill$\blacksquare$

Because $\theta_{\text{\rm sec}}<\theta_{\rm max}$, if (\ref{eqn:innuperp}) holds then the convergence speed is faster than $\theta_{\rm max}$ by Lemma \ref{lem:thetasec}. Next lemma gives a necessary and sufficient condition for guaranteeing (\ref{eqn:innuperp}). 
\begin{lemma}
\label{lem:migikoyuuvector}
A necessary and sufficient condition for $\bm\mu J(\bm\lambda^\ast)\in\bm\nu_{\rm max}^\perp$ to hold for any $\bm\mu\in\bm\nu_{\rm max}^\perp$ is that ${^t}\bm\nu_{\rm max}$ is a right eigenvector for $\theta_{\rm max}$.
\end{lemma}
{\bf Proof:} See Appendix \ref{sec:proofofmigikoyuuvector}.\hfill$\blacksquare$
\begin{figure}[t]
\begin{center}
%\noindent
\begin{overpic}[width=8.8cm]{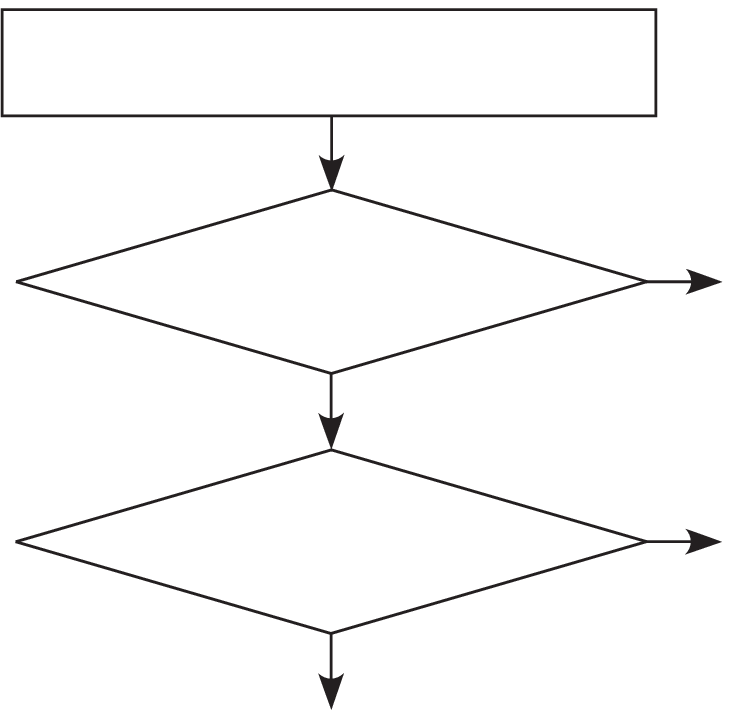}
%\begin{overpic}[width=7.8cm,grid]{./figure/chart.eps}
\put(8,91){Calculate the left eigenvector $\bm\nu_{\rm max}$ for}
\put(8,86){the maximum eigenvalue $\theta_{\rm max}$ of $J(\bm\lambda^\ast)$.}
\put(22,60.5){Is ${^t}\bm\nu_{\rm max}$ a right eigenvector}
\put(22,55.5){for $\theta_{\rm max}$?}
\put(25,25){Does the initial vector}
\put(25,20){$\bm\mu^0$ satisfy $\bm\mu^0{^t}\bm\nu_{\rm max}=0$?}
\put(90,62){No}
\put(90,26){No}
\put(50,6){Yes}
\put(50,40){Yes}
\put(102,59){$\theta_{\rm max}$}
\put(102,22){$\theta_{\rm max}$}
\put(44,-5){$\theta_{\rm sec}$}
\end{overpic}\\[5mm]
\caption{Flow chart for determining the exponential convergence speed}
\label{fig:9}
\end{center}
\end{figure}

If ${^t}\bm\nu_{\rm max}$ is a right eigenvector, then by Lemma \ref{lem:migikoyuuvector}, any $\bm\mu^0\in\bm\nu_{\rm max}^\perp$ yields (\ref{eqn:innuperp}), hence the convergence becomes faster. We will show in the flow chart in Fig.\ref{fig:9} how the convergence speed depends on the choice of initial vector.

Now, we will evaluate the convergence speed for the initial vectors (\ref{eqn:initialmu0}),\,(\ref{eqn:initialbarmu0}) by applying the flow chart. For $J(\bm\lambda^\ast)$ in (\ref{eqn:example5J}), $\theta_{\rm max}=0.702$ and $\theta_{\rm sec}=0.618$. The left eigenvector for $\theta_{\rm max}$ is $\bm\nu_{\rm max}=(-0.500,0.500,0.000)$. We can confirm that ${^t}\bm\nu_{\rm max}$ is a right eigenvector for $\theta_{\rm max}$ and $\bar{\bm\mu}^0{^t}\bm\nu_{\rm max}=0$, thus in Fig.\ref{fig:9} the answers are Yes-Yes, so we reach $\theta_{\rm sec}$. Then by the solid line in Fig.\ref{fig:8}, for $N=500$, we have
\begin{align}
L(500)=0.489\doteqdot-\log\theta_{\rm sec}=0.481.
\end{align}
On the other hand, we have $\bar{\bar{\bm\mu}}^0{^t}\bm\nu_{\rm max}\neq0$, thus the answers are Yes-No, so we reach $\theta_{\rm max}$. Then by the dotted line, for $N=500$, we have
\begin{align}
L(500)=0.360\doteqdot-\log\theta_{\rm max}=0.353.
\end{align}

Checking Example 4 this way, we can see that $\bm\nu_{\rm max}=(-0.431,-0.431,0.862)$ is a left eigenvector for $\theta_{\rm max}=0.855$, but ${^t}\bm\nu_{\rm max}$ is not a right eigenvector. Thus the answer is No, so we reach $\theta_{\rm max}$ and we have (\ref{eqn:example4speedcomparison}).
\end{example}

\subsection{Case (ii): convergence of the $1/N$ order}
\begin{example}
\label{exa:6}
\rm Consider the channel matrix $\Phi^{(2)}$ of (\ref{eqn:Phi2}), i.e.,
\begin{align}
\Phi^{(2)}=\begin{pmatrix}
\,0.800 & 0.100 & 0.100\,\\
\,0.100 & 0.800 & 0.100\,\\
\,0.300 & 0.300 & 0.400\,
\end{pmatrix}.
\end{align}
We have
\begin{align}
\bm\lambda^\ast&=(0.500,0.500,0.000),\\
Q^\ast&=(0.450,0.450,0.100),\\
J(\bm\lambda^\ast)&=\begin{pmatrix}\,0.228 & -0.228 & 0.000\,\\\,-0.228 & 0.228 & 0.000\,\\\,-0.500 & -0.500 & 1.000\,\end{pmatrix},\\
H_3(\bm\lambda^\ast)&=\begin{pmatrix}\,0.000 & 0.000 & -1.000\,\\0.000 & 0.000 & -1.000\\\,-1.000 & -1.000 & -3.990\,\end{pmatrix}.
\end{align}
The eigenvalues of $J(\bm\lambda^\ast)$ are $(\theta_1,\theta_2,\theta_3)=(0.000,0.456,$ $1.000)$. 

We have $N\bm\mu^N$ for $N=500$ as
\begin{align}
&N\bm\mu^N=(-0.510,-0.510,1.019)\\
&\doteqdot\ds\lim_{N\to\infty}N\bm\mu^N=(-0.503,-0.503,1.005).\label{eqn:rironchiexample6}
\end{align}
(\ref{eqn:rironchiexample6}) is obtained by Theorem \ref{the:7}. See Fig.\ref{fig:10}. We can confirm that $N\bm\mu^N$ for large $N$ is close to the limit value in Theorem \ref{the:7}.

\begin{figure}[t]
\begin{center}
%\noindent
\begin{overpic}[width=8cm]{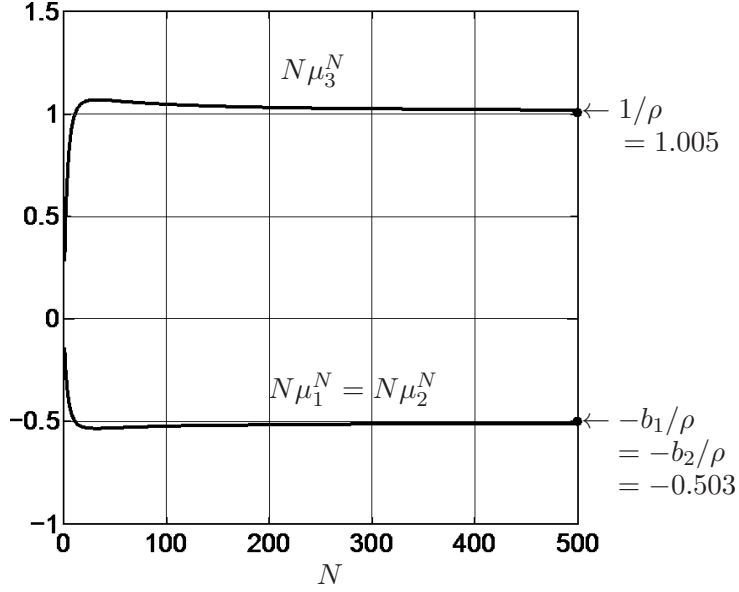}
%\begin{overpic}[width=7cm,grid]{./figure/example6.eps}
\put(95,73.5){$\leftarrow 1/\rho$}
\put(102,68){$=1.005$}
\put(95,22){$\leftarrow$}
\put(101,22){$-b_1/\rho$}
\put(101,16.5){$=-b_2/\rho$}
\put(101,11){$=-0.503$}
\put(51,-4){$N$}
\put(45,80){$N\mu^N_3$}
\put(43,27){$N\mu^N_1=N\mu^N_2$}
\end{overpic}
\caption{Convergence of $N\mu^N_i$ in Example \ref{exa:6}}
\label{fig:10}
\end{center}
\end{figure}
\end{example}

\begin{example}
\label{exa:7}
\rm We will examine another example of slow convergence. Consider the channel matrix
\begin{align}
\Phi^{(5)}\equiv\begin{pmatrix}\,0.720 & 0.215 & 0.065\,\\\,0.013 & 0.431 & 0.556\,\\\,0.250 & 0.700 & 0.050\,\end{pmatrix}.
\end{align}

We have
\begin{align}
\bm\lambda^\ast&=(0.453,0.547,0.000),\\
Q^\ast&=(0.333,0.333,0.334),\\
J(\bm\lambda^\ast)&=\begin{pmatrix}\,0.227 & -0.227 & 0.000\,\\\,-0.188 & 0.188 & 0.000\,\\\,-0.453 & -0.547 & 1.000\,\end{pmatrix},\\
H_3(\bm\lambda^\ast)&=\begin{pmatrix}\,0.000 & 0.000 & -1.000\,\\\,0.000 & 0.000 & -1.000\,\\\,-1.000 & -1.000 & -3.330\,\end{pmatrix}.
\end{align}
The eigenvalues of $J(\bm\lambda^\ast)$ are $(\theta_1,\theta_2,\theta_3)=(0.000,0.416,$ $1.000)$. 

We have $N\bm\mu^N$ for $N=500$ as
\begin{align}
&N\bm\mu^N=(-0.684,-0.825,1.509)\\
&\doteqdot\ds\lim_{N\to\infty}N\bm\mu^N=(-0.682,-0.822,1.504).\label{eqn:rironchiexample7}
\end{align}
See Fig.\ref{fig:11}. We can confirm that $N\bm\mu^N$ for large $N$ is close to the limit value in Theorem \ref{the:7}.
\begin{figure}[t]
\begin{center}
%\noindent
\begin{overpic}[width=8cm]{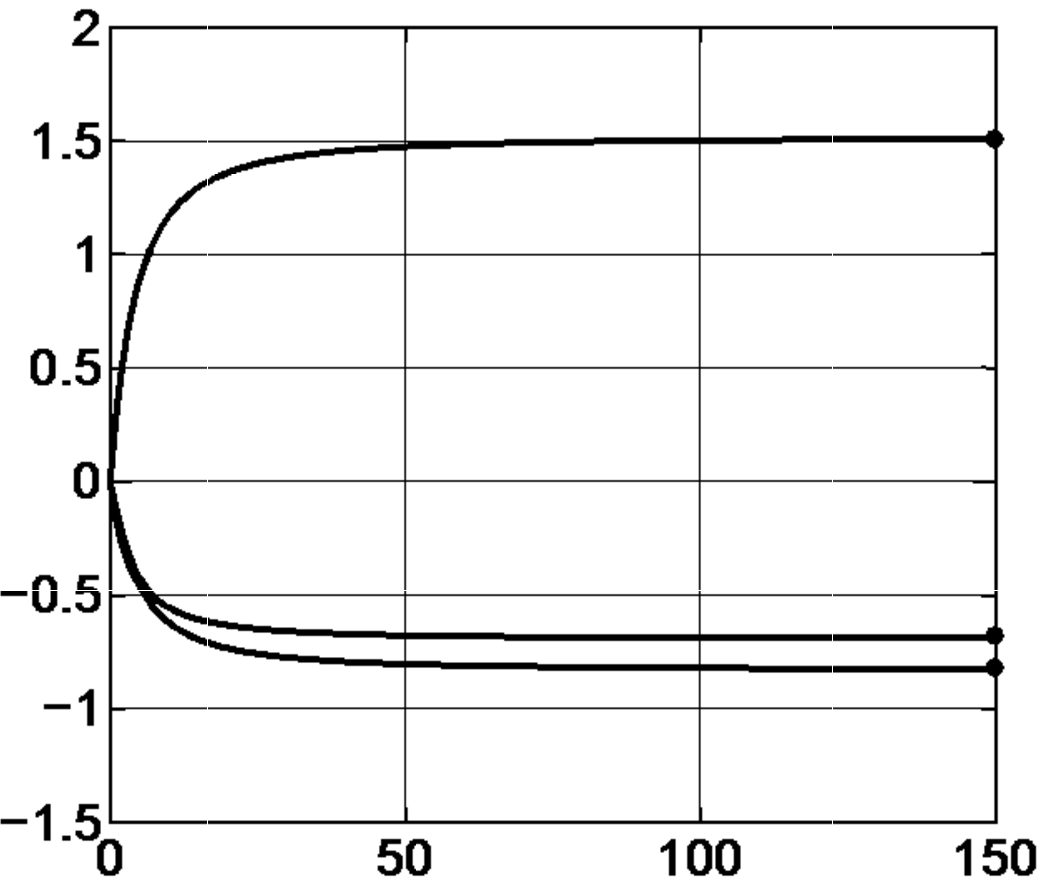}
%\begin{overpic}[width=7cm,grid]{./figure/conv-1.eps}
\put(94.5,69.4){$\leftarrow 1/\rho$}
\put(102,64){$=1.504$}
\put(94.5,23.5){\rotatebox{60}{$\leftarrow$}}
\put(99,29){$-b_1/\rho$}
\put(99.5,23.5){$=-0.682$}
\put(94.1,20.3){\rotatebox{-60}{$\leftarrow$}}
\put(99,14){$-b_2/\rho$}
\put(99,8.5){$=-0.822$}
\put(51,-2){$N$}
\put(45,73){$N\mu^N_3$}
\put(45,27){$N\mu^N_1$}
\put(45,15){$N\mu^N_2$}
\end{overpic}
\caption{Convergence of $N\bm\mu^N$ in Example \ref{exa:7}}
\label{fig:11}
\end{center}
\end{figure}
\end{example}

\subsection{Case (iii): exponential convergence where $\bm\lambda^\ast\in\partial\Delta({\cal X})$}
\begin{example}
\label{exa:8}
\rm Consider the channel matrix $\Phi^{(3)}$ of (\ref{eqn:Phi3}), i.e.,
\begin{align}
\Phi^{(3)}=\begin{pmatrix}
\,0.800 & 0.100 & 0.100\,\\
\,0.100 & 0.800 & 0.100\,\\
\,0.350 & 0.350 & 0.300\,
\end{pmatrix}.
\end{align}
We have
\begin{align}
\bm\lambda^\ast&=(0.500,0.500,0.000),\\
Q^\ast&=(0.450,0.450,0.100),\\
J(\bm\lambda^\ast)&=
\begin{pmatrix}
\,0.228 & -0.228 & 0.000\,\cr
\,-0.228 & 0.228 & 0.000\,\cr
\,-0.428 & -0.428 & 0.856\,\cr
\end{pmatrix}.\label{eqn:example8Jacobimatrix}
\end{align}
The eigenvalues of $J(\bm\lambda^\ast)$ are $(\theta_1,\theta_2,\theta_3)=(0.000,0.456,$ $0.856)$. Then, $\theta_{\rm max}=\theta_3=0.856$. With initial distribution $\bm\lambda^0=(1/3,1/3,1/3)$, we have for $N=500$
\begin{align}
L(500)=0.159\doteqdot-\log\theta_{\rm max}=0.155.
\end{align}
See Fig.\ref{fig:12}.
\begin{figure}[t]
\begin{center}
%\noindent
\begin{overpic}[width=8cm]{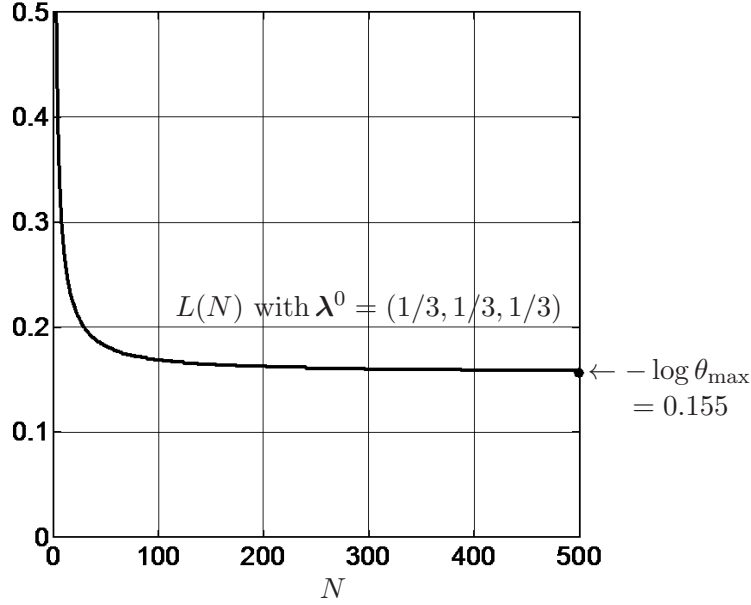}
%\begin{overpic}[width=7cm,grid]{./figure/example8.eps}
\put(95.5,32.5){$\leftarrow-\log\theta_{\rm max}$}
\put(103,26.5){$=0.155$}
\put(51,-4){$N$}
\put(27,43){$L(N)\ \text{\rm with}\,\bm\lambda^0=(1/3,1/3,1/3)$}
\end{overpic}
\caption{Convergence of $L(N)$ in Example \ref{exa:8} with initial distribution $\bm\lambda^0=(1/3,1/3,1/3)$}
\label{fig:12}
\end{center}
\end{figure}

We are here dealing with the exponential convergence in the case (iii) of section \ref{sec:m3narbitray}. In (iii), the Jacobian matrix $J(\bm\lambda^\ast)$ is given by (\ref{eqn:3by3Jacobimatrix3}). Let us consider the (3,3) component $J^{\rm III}=e^{D^\ast_3-C}$ of $J(\bm\lambda^\ast)$ in (\ref{eqn:3by3Jacobimatrix3}) where $0<J^{\rm III}<1$. Putting $\bm{e}_3=(0,0,1)$, we have $J(\bm\lambda^\ast){^t}\bm{e}_3=J^{\rm III}\,{^t}\bm{e}_3$, then $J^{\rm III}$ is an eigenvalue of $J(\bm\lambda^\ast)$ and ${^t}\bm{e}_3$ is a right eigenvector. On the other hand, $\bm{e}_3$ is not a left eigenvector for $J^{\rm III}$. In fact, since every row sum of $J(\bm\lambda^\ast)$ is equal to 0 by Lemma \ref{lem:rowsumofJis0}, putting $\bm{1}=(1,1,1)$, we have $J(\bm\lambda^\ast){^t}\bm{1}=\bm{0}$. Thus, if $\bm{e}_3$ were a left eigenvector for $J^{\rm III}$, then $0=\bm{e}_3J(\bm\lambda^\ast){^t}\bm{1}=J^{\rm III}\bm{e}_3{^t}\bm{1}=J^{\rm III}>0$, a contradiction. Therefore, if $J^{\rm III}=\theta_{\rm max}$, i.e., the maximum of the eigenvalues is achieved in ${\cal I}_{\rm III}$ not in ${\cal I}_{\rm I}$, then by Lemma \ref{lem:migikoyuuvector} or the flow chart in Fig.\ref{fig:9}, we have $L(N)\doteqdot-\log\theta_{\rm max}$ for large $N$. The Jacobian matrix of (\ref{eqn:example8Jacobimatrix}) is one that satisfies $J^{\rm III}=\theta_{\rm max}$.
\end{example}

\section{Conclusion}
In this paper, we investigated the convergence speed of the Arimoto algorithm. First, we noticed that the defining function $F(\bm\lambda)$ of the Arimoto algorithm is a differentiable mapping from the set $\Delta({\cal X})$ of all input distributions into itself. We showed that the capacity achieving input distribution $\bm\lambda^\ast$ is the fixed point of $F(\bm\lambda)$, and analyzed the convergence speed by the Taylor expansion of $F(\bm\lambda)$ about $\bm\lambda=\bm\lambda^\ast$. We concretely calculated the Jacobian matrix $J$ of the first order term of the Taylor expansion and the Hessian matrix $H$ of the second order term. We clarified that if the maximum eigenvalue $\theta_{\rm max}$ of $J(\bm\lambda^\ast)$ satisfies $0\leq\theta_{\rm max}<1$, then the convergence is exponential. Further, we investigated in detail the case that the input alphabet size $m=3$ and the output alphabet size $n$ is arbitrary. We proved, under the assumption $\rho>0$, where $\rho$ was defined in (\ref{eqn:rhodefinition}), the following three conditions are equivalent; type II index in (\ref{eqn:Kuhn-Tucker2}) exists, $\theta_{\rm max}=1$, and the convergence is the $1/N$ order. In this case, we determined the convergence speed by the derivatives of the Kullback-Leibler divergence with respect to the input probabilities. The analysis for the convergence of the $1/N$ order by the Hessian matrix $H$ was done for the first time in this paper.

Based on these analysis, the convergence speeds for several channel matrices were numerically evaluated. As a result, it was confirmed that the convergence speed of the Arimoto algorithm is very accurately approximated by the theoretical values obtained by our theorems.

%\section{Acknowledgement}
%The authors would like to thank Shin-ichiro Hara of Nagaoka University of Technology for giving them the proofs of Lemma \ref{lem:6} and Lemma \ref{lem:migikoyuuvector}, which are key lemmas in this paper. This work was supported by JSPS KAKENHI Grant Number JP17K00008.

%

% or


\newpage

\appendix

\section{Proof of Theorem \ref{the:5}}
\label{sec:proooftheorem4}
{\bf Proof:} Consider the line segment with the start point $\bm\lambda^\ast$ and the end point $\bm\lambda^N$, i.e.,
\begin{align}
\bm\lambda(t)\equiv(1-t)\bm\lambda^\ast+t\bm\lambda^N,\,0\leq t\leq 1.\label{eqn:senbunlstarlN}
\end{align}
The components of (\ref{eqn:senbunlstarlN}) are written by $\lambda_i(t)=(1-t)\lambda^\ast_i+t\lambda^N_i,\,i=1,\cdots,m$. Let us define
\begin{align}
f(t)\equiv F(\bm\lambda(t))\in\Delta({\cal X})
\end{align}
and write its components as $f(t)=(f_1(t),\cdots,f_m(t))$. We have
\begin{align*}
\ds\frac{df_i(t)}{dt}
&=\ds\sum_{i'=1}^m\ds\frac{d\lambda_{i'}(t)}{dt}\left.\ds\frac{\partial F_i}{\partial\lambda_{i'}}\right|_{\bm\lambda=\bm\lambda(t)}\\
&=\ds\sum_{i'=1}^n(\lambda^N_{i'}-\lambda^\ast_{i'})\left.\ds\frac{\partial F_i}{\partial\lambda_{i'}}\right|_{\bm\lambda=\bm\lambda(t)}\\
&=\left((\bm\lambda^N-\bm\lambda^\ast)J(\bm\lambda(t))\right)_i,\,i=1,\cdots,m,
\end{align*}
thus
\begin{align}
\ds\frac{df(t)}{dt}=(\bm\lambda^N-\bm\lambda^\ast)J(\bm\lambda(t)).\label{eqn:dft}
\end{align}

Now, by the relation between the matrix norm and the maximum eigenvalue \cite{hor},\,p.347, for $\epsilon\equiv\theta-\theta_{\rm max}>0$ there exists a vector norm $\|\cdot\|'$ in $\mathbb R^m$ whose associated matrix norm $\|\cdot\|'$ satisfies
\begin{align}
\theta_{\rm max}\leq\|J(\bm\lambda^\ast)\|'<\theta_{\rm max}+\epsilon.
\end{align}
(Note that $'$ does not denote the derivative.) By the continuity of norm, for any $\epsilon_1$ with $0<\epsilon_1<\theta_{\rm max}+\epsilon-\|J(\bm\lambda^\ast)\|'$ there exists $\delta'>0$ such that if $\|\bm\lambda-\bm\lambda^\ast\|'<\delta'$ then $\left|\|J(\bm\lambda)\|'-\|J(\bm\lambda^\ast)\|'\ \right|<\epsilon_1$, especially, $\|J(\bm\lambda)\|'<\|J(\bm\lambda^\ast)\|'+\epsilon_1$. Thus,
\begin{align}
\|J(\bm\lambda)\|'&<\|J(\bm\lambda^\ast)\|'+\theta_{\rm max}+\epsilon-\|J(\bm\lambda^\ast)\|'\\
&=\theta<1.\label{eqn:Jlambdalesstheta}
\end{align}
By the mean value theorem, there exists $t^N\in[0,1]$ which satisfies 
\begin{align}
\|\bm\lambda^{N+1}-\bm\lambda^\ast\|'&=\|F(\bm\lambda^N)-F(\bm\lambda^\ast)\|'\nonumber\\
&=\|f(1)-f(0)\|'\nonumber\\
&\leq\left\|\left.\dfrac{df(t)}{dt}\right|_{t=t^N}\right\|'\,(1-0)\nonumber\\
&=\|(\bm\lambda^N-\bm\lambda^\ast)J(\bm\lambda(t^N))\|'\ \ ({\rm by}\ (\ref{eqn:dft}))\nonumber\\
&\leq\|\bm\lambda^N-\bm\lambda^\ast\|'\,\|J(\bm\lambda(t^N))\|'.\label{eqn:lessnormrecursion}
\end{align}

Here, if $\|\bm\lambda^N-\bm\lambda^\ast\|'<\delta'$ we have $\|J(\bm\lambda^N)\|'<\theta<1$ by (\ref{eqn:Jlambdalesstheta}), so $\|\bm\lambda^{N+1}-\bm\lambda^\ast\|'<\delta'$ by (\ref{eqn:lessnormrecursion}). Thus, by induction, if the initial vector $\bm\lambda^0$ satisfies $\|\bm\lambda^0-\bm\lambda^\ast\|'<\delta'$, then $\|\bm\lambda^N-\bm\lambda^\ast\|'<\delta'$ for all $N$, and so $\|J(\bm\lambda^N)\|'<\theta<1$ by (\ref{eqn:Jlambdalesstheta}).

Therefore by (\ref{eqn:Jlambdalesstheta}),\,(\ref{eqn:lessnormrecursion}), $\|\bm\lambda^{N+1}-\bm\lambda^\ast\|'<\theta\|\bm\lambda^N-\bm\lambda^\ast\|'<\cdots<\theta^{N+1}\|\bm\lambda^0-\bm\lambda^\ast\|'$, so we have
\begin{align}
\|\bm\lambda^N-\bm\lambda^\ast\|'<\theta^N\|\bm\lambda^0-\bm\lambda^\ast\|',\ N=0,1,\cdots.\label{eqn:exponentialdelay}
\end{align}

Finally, we will replace the norm from $\|\cdot\|'$ to the Euclidean norm $\|\cdot\|$. By the equivalence of norms in the finite dimensional vector space \cite{rob}, for the norms $\|\cdot\|'$ and $\|\cdot\|$, there exist constants $K_1>0,\,K_2>0$ such that for arbitrary $\bm\lambda\in\Delta(\cal X)$,
\begin{align}
K_1\|\bm\lambda\|'\leq\|\bm\lambda\|\leq K_2\|\bm\lambda\|'.\label{eqn:normdouchisei}
\end{align}
By (\ref{eqn:exponentialdelay}),\,(\ref{eqn:normdouchisei}),
\begin{align}
\|\bm\lambda^N-\bm\lambda^\ast\|&\leq K_2\|\bm\lambda^N-\bm\lambda^\ast\|'\nonumber\\
&\leq K_2\theta^N\|\bm\lambda^0-\bm\lambda^\ast\|'\nonumber\\
&\leq\ds\frac{K_2}{K_1}\theta^N\|\bm\lambda^0-\bm\lambda^\ast\|,
\end{align}
then putting $K=(K_2/K_1)\|\bm\lambda^0-\bm\lambda^\ast\|,\,\delta=K_1\delta'$, we see that for arbitrary initial vector $\bm\lambda^0$ with $\|\bm\lambda^0-\bm\lambda^\ast\|<\delta$,
\begin{align}
\|\bm\lambda^N-\bm\lambda^\ast\|\leq K\theta^N,\,N=0,1,\cdots
\end{align}
holds.\hfill$\blacksquare$

\section{Proof of Theorem \ref{the:6} (Calculation of Hessian matrix $H_i(\bm\lambda^\ast)$)}
\label{sec:proofoftheorem5}
{\bf Proof:} We will calculate the Hessian matrix $H_i$ of $F_i$ at $\bm\lambda=\bm\lambda^\ast$, i.e., $H_i=(\partial^2F_i/\partial\lambda_{i'}\partial\lambda_{i''}|_{\bm\lambda=\bm\lambda^\ast})$.

Differentiating the both sides of (\ref{eqn:dFi}) by $\lambda_{i''}$, we have
\begin{align}
&\ds\frac{\partial^2F_i}{\partial\lambda_{i'}\partial\lambda_{i''}}\ds\sum_{k=1}^m\lambda_ke^{D_k}+\ds\frac{\partial F_i}{\partial\lambda_{i'}}\ds\frac{\partial}{\partial\lambda_{i''}}\ds\sum_{k=1}^m\lambda_ke^{D_k}\nonumber\\
&\ \ +\ds\frac{\partial F_i}{\partial\lambda_{i''}}\ds\frac{\partial}{\partial\lambda_{i'}}\ds\sum_{k=1}^m\lambda_ke^{D_k}+F_i\ds\frac{\partial^2}{\partial\lambda_{i'}\partial\lambda_{i''}}\ds\sum_{k=1}^m\lambda_ke^{D_k}\nonumber\\
&=\delta_{ii'}e^{D_i}\ds\frac{\partial D_i}{\partial\lambda_{i''}}+\delta_{ii''}e^{D_i}\ds\frac{\partial D_i}{\partial\lambda_{i'}}+\lambda_ie^{D_i}\ds\frac{\partial D_i}{\partial\lambda_{i''}}\ds\frac{\partial D_i}{\partial\lambda_{i'}}\nonumber\\
&\ \ +\lambda_ie^{D_i}\ds\frac{\partial^2D_i}{\partial\lambda_{i'}\partial\lambda_{i''}}.\label{eqn:ddfi}
\end{align}
Here, we will execute the following preliminary calculation.
\begin{align*}
&\ds\frac{\partial^2}{\partial\lambda_{i'}\partial\lambda_{i''}}\ds\sum_{k=1}^m\lambda_ke^{D_k}
=\ds\frac{\partial}{\partial\lambda_{i''}}\left(e^{D_{i'}}+\ds\sum_{k=1}^m\lambda_ke^{D_k}\ds\frac{\partial D_k}{\partial\lambda_{i'}}\right)\\
&=e^{D_{i'}}\ds\frac{\partial D_{i'}}{\partial\lambda_{i''}}+\ds\sum_{k=1}^m\left(\delta_{ki''}e^{D_k}\ds\frac{\partial D_k}{\partial\lambda_{i'}}+\lambda_ke^{D_k}\ds\frac{\partial D_k}{\partial\lambda_{i''}}\ds\frac{\partial D_k}{\partial\lambda_{i'}}\right.\\
&\left.\ \ +\lambda_ke^{D_k}\ds\frac{\partial^2D_k}{\partial\lambda_{i'}\partial\lambda_{i''}}\right)\\
&=e^{D_{i'}}\ds\frac{\partial D_i'}{\partial\lambda_{i''}}+e^{D_{i''}}\ds\frac{\partial D_{i''}}{\partial\lambda_{i'}}+\ds\sum_{k=1}^m\lambda_ke^{D_k}\left(\ds\frac{\partial D_k}{\partial\lambda_{i'}}\ds\frac{\partial D_k}{\partial\lambda_{i''}}\right.\\
&\ \ \left.+\ds\frac{\partial^2D_k}{\partial\lambda_{i'}\partial\lambda_{i''}}\right),
\end{align*}
\begin{align*}
\left.\ds\sum_{k=1}^m\lambda_ke^{D_k}\ds\frac{\partial^2D_k}{\partial\lambda_{i'}\partial\lambda_{i''}}\right|_{\bm\lambda=\bm\lambda^\ast}
&=e^C\ds\sum_{k=1}^{m_1}\lambda_k^\ast\ds\sum_{j=1}^n\ds\frac{P_j^kP_j^{i'}P_j^{i''}}{\left(Q_j^\ast\right)^2}\\
&=e^C\ds\sum_{j=1}^n\ds\frac{P_j^{i'}P_j^{i''}}{Q_j^\ast}\ds\sum_{k=1}^{m_1}\ds\frac{\lambda_k^\ast P_j^k}{Q_j^\ast}\\
&=-e^CD_{i',i''}^\ast,
\end{align*}
\begin{align*}
\left.\ds\frac{\partial^2}{\partial\lambda_{i'}\partial\lambda_{i''}}\ds\sum_{k=1}^m\lambda_ke^{D_k}\right|_{\bm\lambda=\bm\lambda^\ast}
&=e^{D_{i'}^\ast}D_{i',i''}^\ast+e^{D_{i''}^\ast}D_{i'',i'}^\ast\\
&\hspace{-5mm}+e^C\ds\sum_{k=1}^{m_1}\lambda_k^\ast D_{k,i'}^\ast D_{k,i''}^\ast-e^CD_{i',i''}^\ast.
\end{align*}
Based on the above calculation, we substitute $\bm\lambda=\bm\lambda^\ast$ into (\ref{eqn:ddfi}). Define $D_{i,i',i''}^\ast\equiv\partial^2D_i/\partial\lambda_{i'}\partial\lambda_{i''}|_{\bm\lambda=\bm\lambda^\ast}$.
\begin{align*}
&\left.\ds\frac{\partial^2F_i}{\partial\lambda_{i'}\partial\lambda_{i''}}\right|_{\bm\lambda=\bm\lambda^\ast}e^C
=-\left.\ds\frac{\partial F_i}{\partial\lambda_{i'}}\right|_{\bm\lambda=\bm\lambda^\ast}\left(e^{D_{i''}^\ast}-e^C\right)\\
&\ \ -\left.\ds\frac{\partial F_i}{\partial\lambda_{i''}}\right|_{\bm\lambda=\bm\lambda^\ast}\left(e^{D_{i'}^\ast}-e^C\right)\\
&\ \ -F_i^\ast\Big(e^{D_{i'}^\ast}D_{i',i''}^\ast+e^{D_{i''}^\ast}D_{i'',i'}^\ast+e^C\ds\sum_{k=1}^{m_1}\lambda_k^\ast D_{k,i'}^\ast D_{k,i''}^\ast\\
&\ \ -e^CD_{i',i''}^\ast\Big)+\delta_{ii'}e^{D_i^\ast}D_{i,i''}^\ast+\delta_{ii''}e^{D_i^\ast}D_{i,i'}^\ast\\
&\ \ +\lambda_i^\ast e^{D_i^\ast}D_{i,i'}^\ast D_{i,i''}^\ast+\lambda_i^\ast e^{D_i^\ast}D_{i,i',i''}^\ast\\[3mm]
&=e^{D_i^\ast-C}\left\{\delta_{ii'}+\lambda_i^\ast\left(1-e^{D_{i'}-C}+D_{i,i'}^\ast\right)\right\}\left(e^C-e^{D_{i''}^\ast}\right)\\
&\ \ +e^{D_i^\ast-C}\left\{\delta_{ii''}+\lambda_i^\ast\left(1-e^{D_{i''}-C}+D_{i,i''}^\ast\right)\right\}\left(e^C-e^{D_{i'}^\ast}\right)\\
&\ \ +\lambda_i^\ast e^{D_i^\ast-C}\Big(e^CD_{i',i''}^\ast-e^{D_{i'}^\ast}D_{i',i''}^\ast-e^{D_{i''}^\ast}D_{i'',i'}^\ast\\
&\ \ -e^C\ds\sum_{k=1}^{m_1}\lambda_k^\ast D_{k,i'}^\ast D_{k,i''}^\ast\Big)\\
&\ \ +e^{D_i^\ast}\left(\delta_{ii'}D_{i,i''}^\ast+\delta_{ii''}D_{i,i'}^\ast+\lambda_i^\ast D_{i,i'}^\ast D_{i,i''}^\ast+\lambda_i^\ast D_{i,i',i''}^\ast\right).
\end{align*}
By arranging this, we obtain

\medskip

\noindent{\bf Theorem \ref{the:6}}
\begin{align*}
&\left.\ds\frac{\partial^2F_i}{\partial\lambda_{i'}\partial\lambda_{i''}}\right|_{\bm\lambda=\bm\lambda^\ast}=e^{D_i^\ast-C}\Big[(1-e^{D_{i'}^\ast-C}+D_{i,i'}^\ast)(\delta_{ii''}+\lambda_i^\ast(1-e^{D_{i''}^\ast-C}))\\[3mm]
&\ \ +(1-e^{D_{i''}^\ast-C}+D_{i,i''}^\ast)(\delta_{ii'}+\lambda_i^\ast(1-e^{D_{i'}^\ast-C}))\\
&\ \ +\lambda_i^\ast\Big(D_{i,i'}^\ast D_{i,i''}^\ast+D_{i,i',i''}^\ast+D_{i',i''}^\ast-e^{D_{i'}^\ast-C}D_{i',i''}^\ast-e^{D_{i''}^\ast-C}D_{i',i''}^\ast-\ds\sum_{k=1}^{m_1}\lambda_k^\ast D_{k,i'}^\ast D_{k,i''}^\ast\Big)\Big],\\
&\ \ i,i',i''\in{\cal I}.
\end{align*}
Especially, if ${\cal I}_{\rm III}$ is empty, then for $i\in{\cal I}_{\rm II}$,
\begin{align}
&\left.\ds\frac{\partial^2F_i}{\partial\lambda_{i'}\partial\lambda_{i''}}\right|_{\bm\lambda=\bm\lambda^\ast}=\delta_{ii'}D_{i,i''}^\ast+\delta_{ii''}D_{i,i'}^\ast,\,i',i''\in{\cal I},
\end{align}
which is a relatively simple form.\hfill$\blacksquare$
%\end{theorem}
%
\section{Proof of Lemma \ref{lem:thetasec}}
\label{sec:proofofthetasec}
{\bf Proof:} Let $0=\theta_1<\theta_2<\cdots<\theta_{m-1}<\theta_m<1$ be the eigenvalues of $J(\bm\lambda^\ast)$. We have $\theta_{\rm max}=\theta_m$, $\theta_{\rm sec}=\theta_{m-1}$. Let $\bm\nu_i,\,i=1,\cdots,m$, be the left eigenvectors of $J(\bm\lambda^\ast)$ for $\theta_i,\,i=1,\cdots,m$, respectively. We have $\bm\nu_{\rm max}=\bm\nu_m$. Because all the eigenvalues are different, $\{\bm\nu_i\}_{i=1,\cdots,m}$ forms a basis of $\mathbb{R}^m$. Suppose $\bm\mu^N\in\bm\nu_{\rm max}^\perp,\,N=0,1,\cdots$, then $\bm\mu^N$ is uniquely represented as
\begin{align}
\bm\mu^N=\ds\sum_{i=1}^{m-1}\alpha^N_i\bm\nu_i,\,\alpha^N_i\in{\mathbb R},\label{eqn:linearcombination}
\end{align}
in the $m-1$ dimensional subspace $\bm\nu_{\rm max}^\perp$. By (\ref{eqn:linearcombination}), we have
\begin{align}
\bm\mu^{N+1}&=\bm\mu^NJ(\bm\lambda^\ast)\\
&=\ds\sum_{i=1}^{m-1}\alpha^N_i\bm\nu_iJ(\bm\lambda^\ast)\\
&=\ds\sum_{i=1}^{m-1}\alpha^N_i\theta_i\bm\nu_i.\label{eqn:N+1coefficients}
\end{align}
Comparing the coefficients of $\bm\mu^{N+1}=\sum_{i=1}^{m-1}\alpha^{N+1}_i\bm\nu_i$ and (\ref{eqn:N+1coefficients}), we have $\alpha^{N+1}_i=\theta_i\alpha^N_i=\cdots=\left(\theta_i\right)^{N+1}\alpha^0_i,\,i=1,\cdots,m-1$, thus $\bm\mu^N=\sum_{i=1}^{m-1}\left(\theta_i\right)^N\alpha^0_i\bm\nu_i$. Therefore,
\begin{align}
\|\bm\mu^N\|&\leq\ds\sum_{i=1}^{m-1}\left(\theta_i\right)^N|\alpha^0_i|\|\bm\nu_i\|\label{eqn:11}\\
&\leq K\left(\theta_{m-1}\right)^N\label{eqn:12}\\
&=K\left(\theta_{\rm sec}\right)^N,\ K>0.\label{eqn:13}
\end{align}
\hfill$\blacksquare$
\section{Proof of Lemma \ref{lem:migikoyuuvector}}
\label{sec:proofofmigikoyuuvector}
{\bf Proof:} Suppose ${^t}\bm\nu_{\rm max}$ is a right eigenvector for $\theta_{\rm max}$. For any $\bm\mu\in\bm\nu_{\rm max}^\perp$, $\bm\mu J(\bm\lambda^\ast){^t}\bm\nu_{\rm max}=\theta_{\rm max}\bm\mu{^t}\bm\nu_{\rm max}=0$, thus we obtain $\bm\mu J(\bm\lambda^\ast)\in\bm\nu_{\rm max}^\perp$.

Conversely, suppose $\bm\mu J(\bm\lambda^\ast)\in\bm\nu_{\rm max}^\perp$ for any $\bm\mu\in\bm\nu_{\rm max}^\perp$. Our goal is to show $J(\bm\lambda^\ast){^t}\bm\nu_{\rm max}=\theta_{\rm max}{^t}\bm\nu_{\rm max}$, which is equivalent to 
\begin{align}
\bm\mu J(\bm\lambda^\ast){^t}\bm\nu_{\rm max}=\theta_{\rm max}\bm\mu{^t}\bm\nu_{\rm max}\ {\rm holds\ for\ any\ }\bm\mu.\label{eqn:migikoyuuchiequivalent}
\end{align}
We will prove (\ref{eqn:migikoyuuchiequivalent}). Since we can write $\bm\mu$ uniquely as $\bm\mu=K\bm\nu_{\rm max}+\tilde{\bm\mu},\,\tilde{\bm\mu}\in\bm\nu_{\rm max}^\perp$ with constant $K$, we have
\begin{align*}
\bm\mu J(\bm\lambda^\ast){^t}\bm\nu_{\rm max}&=K\bm\nu_{\rm max} J(\bm\lambda^\ast){^t}\bm\nu_{\rm max}+\tilde{\bm\mu} J(\bm\lambda^\ast){^t}\bm\nu_{\rm max}\\
&\hspace{-10mm}=K\theta_{\rm max}\bm\nu_{\rm max}{^t}\bm\nu_{\rm max}+0\ \ ({\rm by\ the\ assumption})\\
&\hspace{-10mm}=\theta_{\rm max}K\bm\nu_{\rm max}{^t}\bm\nu_{\rm max}+\theta_{\rm max}\tilde{\bm\mu}{^t}\bm\nu_{\rm max}\ \ ({\text{\rm by}}\ \tilde{\bm\mu}\in\bm\nu_{\rm max}^\perp)\\
&\hspace{-10mm}=\theta_{\rm max}\bm\mu{^t}\bm\nu_{\rm max},
\end{align*}
which proves (\ref{eqn:migikoyuuchiequivalent}).\hfill$\blacksquare$

\newpage

\begin{thebibliography}{1}
\baselineskip 3mm
%\vspace*{1mm}
%
\bibitem{ahl}
L.V.~Ahlfors, {\it Complex Analysis} (third edition), McGraw-Hill, 1979.
%
\bibitem{ama}
S.~Amari and H.~Nagaoka, {\it Methods of Information Geometry}, American Mathematical Society and Oxford University Press, 2000.
%
\bibitem{ama2}
S.~Amari, {\it Information Geometry and Its Applications}, Applied Mathematical Sciences vol.~194, Springer Japan, 2016.%
\bibitem{ari}
S.~Arimoto, ``An algorithm for computing the capacity of arbitrary discrete memoryless channels'', IEEE Trans.~Inf.~Theory, vol.~18, pp.14-20, Jan. 1972.
%
\bibitem{bla}
R.~E.~Blahut, ``Computation of channel capacity and rate-distortion functions'', IEEE Trans.~Inf.~Theory, vol.~18, no.~4, pp.460-473, Jul. 1972.
%
%\bibitem{ber}
%D.~S.~Bernstein, {\it Matrix Mathematics, Theory, Facts, and Formulas with Application to Linear Systems Theory}, Princ%eton University Press, 2005.
%
%\bibitem{che}
%D.~Cheng, X.~Hu, and C.~Martin, ``On the Smallest Enclosing Balls'', Communications in Information and Systems, vol.~6, no.~2, pp.137-160, 2006.
%
\bibitem{cov}
T.~Cover and J.~Thomas, {\it Elements of Information Theory}, Wiley, June 2006.
%
\bibitem{csi1}
I.~Csisz\`{a}r and J.~K\"{o}rner, {\it Information Theory: Coding Theorems for Discrete Memoryless Systems}, Academic Press, Orlando, 1982.
%
\bibitem{csi2}
I.~Csisz\`{a}r and G.~Tusn\`{a}dy, ``Information Geometry and Alternating Minimization Procedures'', Statistics and Decisions, Supplement Issue No.1, 205-237, 1984.
%
%\bibitem{gal}
%G.~Gallager, {\it Information Theory and Reliable Communication}, Wiley, New York, 1968.
%
\bibitem{hor}R.~A.~Horn and C.~R.~Johnson, {\it Matrix Analysis}, Cambridge University Press, Second Ed., 1985.
%
\bibitem{mat}G.~Matz and P.~Duhamel, ``Information Geometric Formulation and Interpretation of
Accelerated Blahut-Arimoto-Type Algorithms'', in Proceedings of ITW2004, 2004.
%
%\bibitem{mur}
%S.~Muroga, ``On the capacity of a discrete channel.\,I'', J. Phys. Soc. Japan, vol.~8, pp.484-494, 1953.
%
%\bibitem{nag}
%H.~Nagaoka, ``Algorithms of Arimoto-Blahut type for computing quantum channel capacity'', in proceedings of IEEE International Symposium on Information Theory, 1998.
%
\bibitem{nai}
I.~Naiss and H.~Permuter, ``Extension of the Blahut-Arimoto Algorithm for Maximizing Directed Information'', IEEE Trans.~Inf.~Theory, vol.~59, no.~1, pp.204-222, Jan. 2013.
%
\bibitem{naj}
Z.~Naja, F.~Alberge, and P.~Duhamel, ``Geometrical interpretation and improvements of the Blahut-Arimoto's algorithm'', in proceedings of IEEE International Conference on Acoustics, Speech and Signal Processing (ICASSP), 2009.
%
\bibitem{nak1}K.~Nakagawa and F.~Kanaya, ``A New Geometric Capacity Characterization of a Discrete Memoryless Channel'', IEEE Trans.~Inf.~Theory, vol.~34, no.~2, pp.318-321, 1988.
%
\bibitem{nak2}K.~Nakagawa, K.~Watabe and T.~Sabu, ``On the Search Algorithm for the Output Distribution that Achieves the Channel Capacity'', IEEE Trans.~Inf.~Theory, vol.~63, no.~2, pp.1043-1062, Feb. 2017.
%
%
%\bibitem{nak3}K.~Nakagawa and Y.~Takei, ``On the Convergence Speed of the Arimoto Algorithm at the Boundary'', poster presentation at SITA2017, Nov. \color{red}2017\color{black}. (in Japanese)
%
\bibitem{rez}M.~ Rezaeian and A.~Grant, ``A Generalization of Arimoto-Blahut Algorithm'', in proceedings of IEEE ISIT 2004.
%
\bibitem{rob}C.~Robinson, {\it Dynamical Systems: Stability, Symbolic Dynamics, and Chaos}, CRC Press, 1998.
%
%\bibitem{sch}
%S.~Sch\"{o}nherr, ``Quadratic Programming in Geometric Optimization: Theory, Im%plementation, and Applications'', Dissertation of Swiss Federal Institute of Te%chnology, http://www.inf.ethz.ch/personal/emo/DoctThesisFiles/schoenherr02.pdf
%
%
%\bibitem{str}
%G.~Strang, {\it Linear algebra and its applications}, Academic Press, New York, 1976.
%
%\bibitem{val}
%Valentine,\,F.A., {\it Convex Sets}, McGraw-Hill, 1964.
%
%
\bibitem{von}
P.~O.~Vontobel, A.~Ka\v{v}ci\'{c}, D.~M.~Arnold, H.-A.~Loeliger, ``A Generalization of the Blahut-Arimoto Algorithm to Finite-State Channels'', IEEE Trans.~Inf.~Theory, vol.~54, no.~5, pp.1887-1918, 2008.
%
%
\bibitem{yu}
Y.~Yu, ``Squeezing the Arimoto-Blahut Algorithm for Faster Convergence'', IEEE Trans.~Inf.~Theory, vol.~56, no.~7, pp.3149-3157, Jul. 2010.
%
\end{thebibliography}
\end{document}